\documentclass[11pt,reqno]{amsart}
\usepackage{amsaddr}
\usepackage[left=2.5cm,right=2.5cm,top=2cm,bottom=2.5cm]{geometry}
\usepackage{comment}

\usepackage[usenames,dvipsnames]{xcolor}
\usepackage{colortbl}
\usepackage[shortlabels]{enumitem}
\usepackage{amsmath,amsthm,amssymb,bm,bbm,dsfont}
\usepackage{thmtools}
\usepackage{thm-restate}
\usepackage{mathtools}\mathtoolsset{centercolon}

\usepackage[mathscr]{eucal}
\usepackage{upgreek}
\usepackage{setspace}
\usepackage{graphicx}
\usepackage{verbatim}
\usepackage{float}
\usepackage{placeins}
\usepackage{array}
\usepackage{booktabs}

\usepackage{threeparttable}
\usepackage[update,prepend]{epstopdf}
\usepackage{multirow}
\usepackage{amsfonts,amssymb,dsfont}
\usepackage[abs]{overpic}

\usepackage{tikz}
\usepackage{tikzscale}
\usepackage{pgfplots}
\pgfplotsset{compat=1.13}
\usetikzlibrary{calc}
\usetikzlibrary{patterns}
\usetikzlibrary{decorations.pathreplacing}
\usepackage{blox}

\bXLineStyle{-}

\definecolor{sccolor}{RGB}{62, 150, 81}
\definecolor{spcolor}{RGB}{57, 106, 177}

\definecolor{dullmagenta}{rgb}{0.4,0,0.4}   
\definecolor{darkblue}{rgb}{0,0,0.4}
\usepackage{hyperref}
\hypersetup{
	unicode=false,          
	pdftoolbar=true,        
	pdfmenubar=true,        
	pdffitwindow=false,     
	pdfstartview={FitH},    
	pdftitle={My title},    
	pdfauthor={Author},     
	pdfsubject={Subject},   
	pdfcreator={Creator},   
	pdfproducer={Producer}, 
	pdfkeywords={keyword1} {key2} {key3}, 
	pdfnewwindow=true,      
	linktocpage=true,
	colorlinks=true,        
	linkcolor=Red,          
	citecolor=ForestGreen,  
	filecolor=Magenta,      
	urlcolor=BlueViolet,    
}
\usepackage{doi}
\usepackage{caption, subcaption}
\usepackage{enumitem}
\usepackage{shadethm}

\makeatletter
\newcommand{\opnorm}{\@ifstar\@opnorms\@opnorm}
\newcommand{\@opnorms}[1]{%
	\left|\mkern-1.5mu\left|\mkern-1.5mu\left|
	#1
	\right|\mkern-1.5mu\right|\mkern-1.5mu\right|
}
\newcommand{\@opnorm}[2][]{%
	\mathopen{#1|\mkern-1.5mu#1|\mkern-1.5mu#1|}
	#2
	\mathclose{#1|\mkern-1.5mu#1|\mkern-1.5mu#1|}
}
\makeatother

\makeatletter
\ifx\@NODS\undefined%

\else%
\fi
\makeatother

\makeatletter
\ifx\@NODS\undefined%

\let\mathbb=\mathds
\else%
\fi
\makeatother

\usepackage[backend=bibtex,
hyperref=true,
url=true,
isbn=true,
backref=false,
style=ieee,
sorting=none,
block=none]{biblatex}
\bibliography{reference.bib}

\AtEveryBibitem{
    \clearfield{urlyear}
    \clearfield{urlmonth}
}
\newbibmacro{string+doi}[1]{%
	\iffieldundef{doi}{#1}{\href{http://dx.doi.org/\thefield{doi}}{#1}}}
\DeclareFieldFormat{title}{\usebibmacro{string+doi}{\mkbibemph{#1}}}
\DeclareFieldFormat[journal]{title}{\usebibmacro{string+doi}{\mkbibquote{#1}}}

\renewbibmacro*{issue+date}{%
	\iffieldundef{issue}
	{}
	{\printtext[parens]{\printfield{issue}}}%
	\newunit}%
\DeclareFieldFormat{doilink}{
	\iffieldundef{doi}{#1}
	{\href{http://dx.doi.org/\thefield{doi}}{#1}}}

\DeclareBibliographyDriver{article}{%
	\usebibmacro{bibindex}%
	\usebibmacro{begentry}%
	\usebibmacro{author/translator+others}%
	\setunit{\labelnamepunct}\newblock
	\usebibmacro{title}%
	\newunit
	\printlist{language}%
	\newunit\newblock
	\usebibmacro{byauthor}%
	\newunit\newblock
	\usebibmacro{bytranslator+others}%
	\newunit\newblock
	\printfield{version}%
	\newunit\newblock
\iffieldundef{doi}{%
	\usebibmacro{journal+issuetitle}%
		\newunit\newblock
		\usebibmacro{byeditor+others}%
		\newunit\newblock
		\usebibmacro{pages}
		\newunit\newblock
		\usebibmacro{date}
}{%
	\href{http://dx.doi.org/\thefield{doi}}{%
		\usebibmacro{journal+issuetitle}%
		\newunit\newblock
		\usebibmacro{byeditor+others}%
		\newunit\newblock
		\usebibmacro{pages}
		\newunit\newblock
		\usebibmacro{date}
		}%
		}%
	\newunit\newblock
	\iftoggle{bbx:isbn}
	{\printfield{issn}}
	{}%
	\newunit\newblock
	\usebibmacro{doi+eprint+url}%
	\newunit\newblock
	\usebibmacro{addendum+pubstate}%
	\newunit\newblock
	\usebibmacro{pageref}%
	\usebibmacro{finentry}
}

\usepackage{autonum}

\usepackage{xargs}
\usepackage[prependcaption]{todonotes}
\newcommandx{\eric}[2][1=]{\todo[inline, author={Eric}, linecolor=yellow,backgroundcolor=yellow!25,bordercolor=yellow,#1]{#2}}

\newcommandx{\ericside}[2][1=]{\todo[author={Eric}, linecolor=yellow,backgroundcolor=yellow!25,bordercolor=yellow,#1]{#2}}

\DeclarePairedDelimiter{\ceil}{\lceil}{\rceil}
\DeclarePairedDelimiter{\floor}{\lfloor}{\rfloor}

\DeclareMathOperator{\Tr}{Tr}
\DeclareMathOperator{\tr}{Tr}
\DeclareMathOperator{\e}{\mathrm{e}}

\DeclareMathOperator{\one}{\mathds{1}}

\newcommand{\Pe}{P_\textnormal{e}}
\newcommand{\Peavg}{P_\textnormal{e}\!^\textnormal{avg}}
\newcommand{\Pemax}{P_\textnormal{e}\!^\textnormal{max}}

\newcommand{\Pestarchan}{P^{\star}_\textnormal{e, c}}
\newcommand{\Pestarsource}{P^{\star}_\textnormal{e, s}}

\DeclareMathOperator{\rate}{rate}

\newcommand{\source}{_\textnormal{s}}
\newcommand{\chan}{_\textnormal{c}}

\newcommand{\rs}{_\textnormal{r\!,s}^\downarrow}
\newcommand{\rc}{_\textnormal{r\!,c}^\downarrow}

\newcommand{\sps}{_\textnormal{sp,s}}
\newcommand{\spc}{_\textnormal{sp,c}}

\newcommand{\scs}{_\textnormal{sc,s}^*}
\newcommand{\scc}{_\textnormal{sc,c}^*}

\newcommand\restr[2]{{
  \left.\kern-\nulldelimiterspace 
  #1 
  \vphantom{\big|} 
  \right|_{#2} 
  }}

\newcommand{\inv}{^{-1}}

\newcommand{\bra}[1]{\langle #1 |}
\newcommand{\ket}[1]{| #1 \rangle}

\newcommand{\be}{{\mathbf e}}

\newcommand{\bu}{{\mathbf u}}

\newcommand{\bx}{\mathbf{x}}

\newcommand{\by}{{\mathbf y}}

\newcommand{\cH}{{\mathcal{H}}}

\newcommand{\cM}{\mathcal{M}}

\newcommand{\cE}{{\mathcal{E}}}
\newcommand{\sC}{{\mathcal{C}}}
\newcommand{\cD}{{\mathcal{D}}}
\newcommand{\cC}{{\mathcal{C}}}

\newcommand{\cX}{{\mathcal{X}}}
\newcommand{\cZ}{{\mathcal{Z}}}

\newcommand{\cs}{{\mathcal{Z}}}

\def\0{{\mathbf{0}}}
\def\1{{\mathbf{1}}}
\def\2{{\mathbf{2}}}
\def\3{{\mathbf{3}}}
\def\4{{\mathbf{4}}}
\def\5{{\mathbf{5}}}
\def\6{{\mathbf{6}}}

\def\7{{\mathbf{7}}}
\def\8{{\mathbf{8}}}
\def\9{{\mathbf{9}}}

\def\be{\begin{equation}}
\def\ee{\end{equation}}
\def\bea{\begin{eqnarray}}
\def\eea{\end{eqnarray}}

\makeatletter
\def\thmheadbrackets#1#2#3{%
  \thmname{#1}\thmnumber{\@ifnotempty{#1}{ }\@upn{#2}}%
  \thmnote{ {\the\thm@notefont[#3]}}}
\makeatother

\newtheoremstyle{brackets}
  {}
  {}
  {\itshape}
  {}
  {\bfseries}
  {.}
  { }
  {\thmheadbrackets{#1}{#2}{#3}}

  \theoremstyle{brackets}
\newtheorem*{theo*}{Theorem}

\theoremstyle{plain}
\newtheorem{prop}{Proposition}[section] 
\newtheorem{theo}[prop]{Theorem} 

\newtheorem{lemm}[prop]{Lemma} 
\newshadetheorem{conj}{Conjecture}

\theoremstyle{definition}

\theoremstyle{remark}
\newtheorem{remark}[prop]{Remark}

\begin{document}
	
\let\origmaketitle\maketitle
\def\maketitle{
	\begingroup
	\def\uppercasenonmath##1{} 
	\let\MakeUppercase\relax 
	\origmaketitle
	\endgroup
}

\title{\bfseries \Large{Duality between source coding with quantum side information and c-q channel coding  }}

\author{ {Hao-Chung Cheng$^{1,2}$, Eric P. Hanson$^{3}$, Nilanjana Datta$^{3}$, Min-Hsiu Hsieh$^1$ }}
\address{\small  	
	$^{1}$University of Technology Sydney, Australia\\
	$^{2}$National Taiwan University, Taiwan (R.O.C.)\\
	$^{3}$Cambridge University, United Kingdom}
 
\date{\today}

\begin{abstract}
In this paper, we establish an interesting duality between two different quantum information-processing tasks, namely, classical source coding with quantum side information, and channel coding over c-q channels. The duality relates the optimal error exponents of these two tasks, generalizing the classical results of Ahlswede and Dueck. We establish duality both at the operational level and at the level of the entropic quantities characterizing these exponents. For the latter, the duality is given by an exact relation, whereas for the former, duality manifests itself in the following sense: an optimal coding strategy for one task can be used to construct an optimal coding strategy for the other task. Along the way, we derive a bound on the error exponent for c-q channel coding with constant composition codes which might be of independent interest.
\end{abstract}
\maketitle

\section{Introduction} \label{sec:introduction}

Duality is one of the most elegant and useful notions in mathematics. It allows one to relate pairs of seemingly different concepts, mathematical structures, problems and their solutions. In the words of Atiyah \cite{MA}, duality is a ``principle", the first consideration of which dates back many centuries. It has found applications in diverse fields of mathematics, ranging from group theory, topology, analysis and geometry to convex optimization and information theory. In physics, duality plays a fundamental role in fields like quantum mechanics, electromagnetism and quantum field theory. Over the years, the notion of duality has been generalized, adapted and modified in various different contexts.

The first observation of a ``curious and provocative" duality in information theory was by Shannon \cite{Sha59}. He pointed out that the fundamental information-theoretic problems of data compression (or source coding) and data transmission (or channel coding) can be studied as information-theoretic duals of each other. More precisely, this duality is evident for the following pair of source- and channel coding  problems: (a) lossy data compression (or rate distortion), in which the decompressed data is required to satisfy a certain distortion constraint, and (b) channel coding with cost constraint, in which one associates a cost function to the channel inputs. In the former, the aim is to find the optimal stochastic map between the input source alphabet and the output alphabet of the compression-decompression scheme, which {\em{minimizes}} the mutual information between the input and output of the scheme. In the latter, the aim is to find the optimal input distribution for the given channel (i.e.~stochastic map) which {\em{maximizes}} the mutual information between the channel input and output, subject to the cost constraint. This duality was exploited to design iterative algorithms for the optimization of mutual information \cite{Bla72, Ari72}, and later Gupta and Verd\'u studied under what circumstances the capacity-achieving channel encoder-decoder pair attains the rate-distortion function \cite{GV11}.

Pradhan {\em{et al}}, \cite{PCR03} described a different duality between source- and channel coding as follows. They specified the conditions under which a pair of encoding and decoding maps for source coding (respectively, channel coding) is a functional dual to a channel (respectively, source) coding scheme, in the sense that the optimal encoding map for one problem is functionally identical to the optimal decoding
map for the other problem. The notion of duality was also extended to source and channel coding with side information \cite{CC02, PCR03, WV04} and to  multiuser systems \cite{PR06}.

Moreover, the particular case of source coding with side information at the decoder (so-called Slepian-Wolf coding \cite{SW73}) exhibits a natural duality with channel coding. Letting $X$ and $Y$ denote the random variables of the source sequences and the side information, respectively, the conditional probability $\{P_{Y|X}(Y|x)\}_x$ can be interpreted as giving the transition probabilities describing a channel. Csisz{\'a}r recognized this relation \cite{Csi82} and pointed out a mirror-symmetry between the entropic error exponents of Slepian-Wolf coding and channel coding \cite{SGB67, SGB67b, Gal65, Har68, Gal68, Gal72, Kos77, CK80, CK81, Csi82, CK11}. Moreover, if the induced channel satisfies a so-called cyclic symmetry condition, an identity between the entropic error exponents between source and channel coding can be proved. It was later shown that certain linear block codes that cause a channel decoding error also leads to a Slepian-Wolf decoding error, and vice versa \cite{CHJ+09}. In addition, Ahlswede and Dueck showed how to construct source codes with side information from channel codes and vice-versa, and used this construction to establish a duality between the associated asymptotic entropic error exponents \cite[Theorem 1]{AD82}.

Although the dualities in classical information theory have been extensively studied, they have not been studied much in the quantum regime.  Inspired by the classical setting \cite{SV94, SV96, Win02}, it was shown that the channel simulation with quantum side information can be used to construct a lossy data compression protocol \cite{LD09, DRRW13, DHW13}. Additionally, Renes considered classical data compression with quantum side information in which an initial joint quantum state was partially measured to yield the classical random variable to be compressed. {He showed that in this case, such a data compression protocol can be used to implement privacy amplification against a quantum adversary for a classical random variable resulting from performing a certain complementary measurement \cite{Ren10}}.  Renes also considered a notion of duality via complementary channels to relate lossy source coding with privacy amplification \cite{Ren17}, and channel coding with randomness extraction \cite{Ren18}.            

In this paper, we consider a generalization of the duality studied by Ahlswede and Dueck, namely a duality between the optimal error exponent for classical source coding with quantum side information and classical-quantum channel coding. This stems from a construction of source codes from channel codes and vice-versa, and hence can be termed \emph{operational duality}.

We consider an $n$-blocklength classical-quantum (c-q) source that outputs sequences $\bx$ with values in a finite alphabet $\cX^n$ with some probability $p(\bx)$, along with a quantum state $\rho_{B^n}^{\bx}$ on a Hilbert space $\cH_{B}^{\otimes n}$. When the sequence is $\bx = (x_1,\dotsc,x_n)$, the corresponding quantum state is $\rho_{B^n}^{\bx} = \rho_B^{x_1}\otimes \dotsm \otimes \rho_B^{x_n}$. Such a source can be characterized by a c-q state
\[
\rho_{X^nB^n} = \sum_{\bx\in \cX^n} p(\bx)\ket{\bx}\bra{\bx}\otimes \rho_{B^n}^{\bx}.
\]
In a compression-decompression scheme, this quantum state  plays the role of \emph{quantum side information} and is sent to the decoder.  In a fixed-length compression-decompression scheme for a c-q source, each sequence $\bx \in \cX^n$ is mapped to a shorter sequence in a set $\cZ$  with size $|\cZ| = \ceil{2^{nR}}$ by an encoding map $\cE\source$. Here, $R\in(0, \log |\cX|)$ is the rate of the compression-decompression scheme; $R$ bits are used per letter after compression. To decompress, the compressed sequence $\cE(\bx)$ along with the quantum state $\rho_{B^n}^\bx$ is subject to a decoding map $\cD\source$, which outputs a sequence $\hat \bx \in \cX^n$. If $\bx = \hat \bx$, then the protocol has successfully compressed and decompressed $\bx$ (with the help of the side information $\rho_{B^n}^{\bx}$ at the decoder); otherwise, an error has occurred. Together, the encoder and decoder constitute a compression-decompression code $(\cE\source,\cD\source)$ of rate $R$ for the source $\rho_{X^nB^n}$.

The simplest example of an $n$-blocklength c-q source is a memoryless (or i.i.d.) one for which $\rho_{X^nB^n} = \rho_{XB}^{\otimes n}$ for some c-q state $\rho_{XB}$. We consider the minimal probability of error over all compression-decompression codes of rate $R$ for $\rho_{XB}^{\otimes n}$, denoted $\Pestarsource(n,R)$, and the associated exponent $e\source(n,R) = -\frac1n \log(\Pestarsource(n,R))$. The subscript $s$ is used to denote source coding, as opposed to channel coding.

Besides i.i.d.~sources, another class of c-q source is of particular relevance to this work, namely the class of uniform sources of a fixed type. Recall that the type of a sequence $\bx \in \cX^n$ is its empirical distribution. For a fixed type $Q$, we consider a c-q source which outputs any sequence $\bx = (x_1,\dotsc,x_n)$ of type $Q$ with uniform probability (over all sequences of type $Q$), along with quantum side information $\rho_B^{x_1}\otimes \dotsm \rho_B^{x_n}$. We denote the c-q state associated to the source by $\rho_{X^nB^n}^{(Q)}$; note that this state only depends on $Q$ and the quantum side information $\{\rho_B^x\}_{x\in \cX}$. Thus, while $\rho_{X^nB^n}^{(Q)}$ is not an i.i.d. source, it is still highly structured. As before, we consider the minimal probability of error over all compression-decompression codes of rate $R$ for this source, denoted $\Pestarsource(n,R,Q)$, and the associated exponent $e\source(n,R,Q) = -\frac1n \log(\Pestarsource(n,R,Q))$.

The i.i.d~source $\rho_{XB}^{\otimes n}$ can be written as a convex combination of the states $\rho_{X^nB^n}^{(Q)}$, where $Q$ ranges over all possible types:
\begin{equation} \label{eq:type-decomp-intro}
\rho_{XB}^{\otimes n} = \sum_Q \Pr( \bx \text{ has type } Q)\, \rho_{X^nB^n}^{(Q)}.
\end{equation}
 This is called the \emph{type decomposition} of the i.i.d.~source, and provides a fundamental relation between $\rho_{X^nB^n}^{(Q)}$ and $\rho_{XB}^{\otimes n}$.

The map $x\mapsto \rho_B^x$ plays a central role in defining the above sources. In fact, the map $\mathscr{W}: \cX\to S(B)$, $\mathscr{W}(x) = \rho_B^x$ is a \emph{c-q channel}, sending classical letters to quantum states. This naturally leads to the consideration of the transmission of classical information via $n$ uses of this channel. Given a set of messages, $\cM$, a message $m\in \cM$ is encoded by a channel encoder $\cE\chan: \cM \to \cX^n$, transmitted through the channel $\mathscr{W}^{\otimes n}$, and decoded by a channel decoder $\cD\chan$ which measures the output state $\mathscr{W}^{\otimes n}\circ \cE (m)$ and yields a message $\hat m\in \cM$. An error occurs when $\hat m\neq m$. The pair $(\cE\chan,\cD\chan)$ defines an $n$-blocklength c-q channel code for $\mathscr{W}$. The code has rate $R$ when $|\cM| = \floor{2^{nR}}$; that is, when $R$ messages can be sent per use of the channel. The average probability of error (assuming each message $m$ is sent uniformly at random) minimized over all c-q channel codes of rate $R$ for the channel $\mathscr{W}$ is denoted $\Pestarchan(n,R)$, and the associated exponent by $e\chan(n,R) = -\frac1n \log(\Pestarchan(n,R))$. Here, the subscript $c$ denotes channel coding.

We consider a restricted class of code for a fixed type $Q$, called a constant composition code of type $Q$, for which the encoder $\cE\chan: \cM\to \cX^n$ is required to only output sequences in $\cX^n$ of type $Q$. The minimal probability of error over all constant composition codes of type $Q$ and rate $R$ for the channel $\mathscr{W}$ is denoted $\Pestarchan(n,R,Q)$, and the associated exponent $e\chan(n,R, Q) = -\frac1n \log(\Pestarchan(n,R,Q))$.

We show that for any type $Q$,
\begin{equation}\label{eq:intro-duality-Q}
e\source(n,R,Q)\approx e\chan(n,H(Q)-R, Q).
\end{equation}
\begin{figure}[ht]
\centering
\includegraphics{combined.tikz}            
\caption{A protocol to compress classical data with quantum side information, in which the quantum side information is realized as a c-q channel $\mathscr{W}: x\mapsto \rho_B^x$, for the source $\rho_{X^nB^n}^{(Q)}$. The source encoder $\cE\source$ compresses the classical messages $\bx$ to a set of size $|\cZ_n| = \ceil{2^{nR}}$. The source decoder $\cD\source$ receives the pair $\rho_{B^n}^{\bx}$ and $\cE\source(\bx)$ and attempts to recover $\bx$. The classical data $X^n$ is uniform over $T_Q^n$, and has entropy approximately $n H(Q)$. The source decoder receives $nR$ bits from the source encoder, and therefore needs to recover approximately $n(H(Q)-R)$ bits from $\mathscr{W}^{\otimes n}$. This intuition is reflected in the duality formula \eqref{eq:intro-duality-Q}.}            

\end{figure}
This expression can be understood intuitively by considering a `combined' figure. In the source coding task with quantum side information at the decoder, the quantum side information can be considered to be generated by sending a copy of the classical message $\bx$ through the channel $\mathscr{W}^{\otimes n}$ and supplying the result to the decoder. Since $\rho_{X^nB^n}^{(Q)}$ outputs sequences of type $Q$ with uniform probability, its entropy is approximately $nH(Q)$, and therefore has $nH(Q)$ bits of information associated to it. When compressing this information at a rate $R$, the source decoder receives $nR$ bits of information from the source encoder. The discrepancy, $n(H(Q)-R)$, must therefore be supplied by the side information, and the channel $\mathscr{W}$ must therefore be operated at a rate $H(Q)-R$. The probability of error for this source protocol can therefore be bounded by the minimal probability of error for operating $\mathscr{W}$ at a rate $H(Q)-R$. On the other hand, by the same logic any compression-decompression scheme for the source $\rho_{X^nB^n}^{Q}$ can be seen as implicitly operating the channel $\mathscr{W}$ at a rate $H(Q)-R$, and therefore the probability of error for the channel coding task can be bounded by the optimal probability of error for the source coding task. In order to relate the error exponents, it is important that $\rho_{X^nB^n}^{(Q)}$ is uniform over all sequences of type $Q$  because the probability of error for the c-q channel coding task is defined as an average of the probability of error for each message, each sent with uniform probability.

This duality can be raised to the level of i.i.d.~sources $\rho_{XB}^{\otimes n}$ and c-q channel codes with arbitrary encoders using the type decomposition given in~\eqref{eq:type-decomp-intro}. It can be seen that $\Pr( \bx \text{ has type } Q) \approx \e^{-nD(Q\|p)}$, where $p$ is the probability distribution of the source. Since asymptotically the smallest exponent dominates the others, we find
\[
e\source(n,R) \approx \min_Q [ e\chan(n,H(Q)-R,Q) + D(Q\|p)]
\]
using the duality at the level of fixed types. In \cite{CHDH-2018}, the present authors determined upper and lower bounds on the source coding exponent $e\source(n,R)$ in terms of certain entropic quantities which we collectively denote here as $E\source(R)$. In the present work, we determine entropic bounds on $e\chan(n,H(Q)-R,Q)$. The duality \eqref{eq:intro-duality-Q} therefore yields entropic bounds on $e\source(n,R,Q)$ in terms of certain entropic quantities we denote by $E\source(R,Q)$. Interestingly, we find the following exact relation
\begin{equation}
E\source(R) = \min_Q [ E\source(R,Q) + D(Q\|p)]
\end{equation}
for some of these quantities.

This paper is organized as following. We introduce the protocol and notation of c-q Slepian-Wolf coding and channel coding in Section~\ref{sec:prelim}.
We present our main results of the operational duality in Section~\ref{sec:operational_duality}.
In Section~\ref{sec:entropic_duality}, we prove a new achievability bound for c-q channel coding and show the entropic duality between exponent functions.  Lastly, we conclude this paper in Section~\ref{sec:conclusions}.

\section{The information-theoretic tasks}\label{sec:prelim}

This section introduces the two quantum information-processing tasks, the classical source compression with quantum side information and the channel coding over classical-quantum channels.

\subsection{Classical-Quantum Slepian-Wolf (CQSW) Source Coding}

Consider a classical-quantum (c-q) state 
\begin{equation} \label{eq:rho_XB-cq}
\rho_{X^nB^n} = \sum_{\bx\in \cX^n} p(\bx)\ket{\bx}\bra{\bx}\otimes \rho_{B^n}^\bx
\end{equation}
where $\cX$ is a finite alphabet and $\rho_{B^n}^\bx := \rho_B^{x_1} \otimes \rho_B^{x_2} \otimes \dotsm \otimes \rho_B^{x_n}$ for $\bx=(x_1,\dotsc,x_n)$ is a product quantum state; each $\rho_{B}^{x_j} \in S(B)$, where $S(B)$ is the set of density matrices on a Hilbert space, $\cH_B$. We describe the task of compressing sequences $\bx$, which occur with probability $p(\bx)$, from a classical source such that the decoder has access to $\rho_{B^n}^{\bx}$. We consider a compressed space $\cZ$, which is either a finite set or the set of all finite length binary sequences, $\{0,1\}^*$; an encoding map $\cE: \cX^n \to \cZ$, and a decoding map $\cD : \cZ \times S(B^n) \to \cX^n$, where $S(B^n)$ denotes the set of density matrices on $\cH_{B^n} = \cH_B^{\otimes n}$. If we fix the first argument, the compressed classical message, as some $z\in \cZ$, then the map $\cD(z, \cdot) : S(B^n) \to \cX^n$ is given by a positive operator-valued measurement (POVM). Thus, we can represent the decoding by a collection of POVMs $\{\cD^{(z)}\}_{z\in \cZ}$, where $\cD^{(z)} = \{\Pi_{\bx}^{(z)}\}_{\bx\in \cX^n}$ with $\Pi_{ \bx}^{(z)}\geq 0$ and $\sum_{ \bx\in \cX^n} \Pi_{ \bx}^{(z)} = \one$,  for each $z\in \cZ$.  That is, if Alice sends the message $\bx$, Bob receives $\cE(\bx)$, and measures the state $\rho_{B^n}^\bx$ with the POVM $\{\Pi_{\bx}^{(\cE(\bx))}\}_{\bx\in \cX^n}$.

We may embed source coding with classical side information in this framework.  The side information is \emph{classical} when there is an orthonormal basis $\{\ket{y}\}$ of $\cH_B$ such that for each $x\in \cX$,
\begin{equation} \label{eq:CSI}
\rho_B^x = \sum_y p(y|x) \ket{y}\bra{y}
\end{equation}
for some probability distribution $\{p(y|x)\}_{y}$. In this case, $\rho_{XB}$ can equivalently be described by the joint random variables $(X,Y)$ which have distribution $\Pr(X=x, Y=y) = p(x)p(y|x)$. Note that in this case, $\dim \cH_B$ is the size of the alphabet associated to the random variable $Y$.

A CQSW code $\cC$ for the state $\rho_{X^nB^n}$ is a tuple $\cC = (\cE,\cD)$ consisting of an encoding map and a decoding map.
For such a code, the average probability of error is given by
\begin{equation} \label{eq:Pe_source}
\Pe(\sC)\equiv  \Peavg(\rho_{XB},\sC) = 1 - \sum_{\bx \in \cX^n} p(\bx) \tr[ \Pi_{\bx}^{(\cE(\bx))} \rho_{B^n}^{\bx} ].
\end{equation} 
Similarly, the maximal probability of error is given by
\begin{equation}
 \Pemax(\rho_{XB},\sC) = \max_{\bx\in \cX^n}(1 -  \tr[ \Pi_{\bx}^{(\cE(\bx))} \rho_{B^n}^{\bx} ]).
\end{equation}
We will only consider the average probability of error for source coding, however.

In the previous discussion, we have only considered deterministic encoding maps; in fact, this is without loss of generality. Consider a random encoding which maps $\bx$ to $z\in \cZ$ with some probability $P(z|\bx)$. In this case, the probability of error is given by
\begin{equation}
\Pe(\sC) = 1 - \sum_{\substack{\bx \in \cX^n\!,\\z \in \cZ}} P(z|\bx) p(\bx) \tr[ \Pi_{\bx}^{(z)} \rho_{B^n}^{\bx} ].
\end{equation}
Alternatively, we can see the random encoding $\cE$ as applying a deterministic encoding $\cE_j$ with some probability $Q_j$. Then for a code $\sC = (\cE, \cD)$,
\begin{equation} \label{eq:Pe_rand_as_det}
\Pe((\cE, \cD)) = 1 - \sum_j Q_j \sum_{\bx \in \cX^n}  p(\bx) \tr[ \Pi_{\bx}^{(\cE_j(\bx))} \rho_{B^n}^{\bx} ] = \sum_j Q_j \Pe((\cE_j,\cD)).
\end{equation}
Thus, the error probability for a random encoding is an average of error probabilities of deterministic encodings. In particular, $\min_j \Pe((\cE_j, \cD)) \leq \Pe((\cE, \cD))$, so the optimal error probability is achieved for a deterministic code. Therefore, we will restrict to deterministic encodings in this work.

We consider \emph{fixed length} coding in which each sequence $\bx$ is mapped to a shorter sequence of fixed length by the encoding map, i.e. $|\cZ| = \ceil{2^{nR}}$ for some $R \in (0,\log |\cX|)$ which is called the \emph{rate}. Without loss of generality we assume surjectivity of the encoding map, namely $\cE(\cX^n) = \cZ$.

We consider the two following forms for the probability distribution $p(\bx)$ on $\cX^n$.

\begin{enumerate}[(i)]
 	\item \textbf{The i.i.d. case:} 
	 $p(\bx) = p(x_1)\dotsm p(x_n)$ for some probability distribution $p$ on $\cX$. Without loss of generality, we assume $p(x)> 0$ for each $x\in \cX$. In this case, \eqref{eq:rho_XB-cq}
 becomes
\begin{equation}
\rho_{X^nB^n} = \rho_{XB}^{\otimes n} = \left( \sum_{x\in \cX} p(x) \ket{x}\bra{x}\otimes \rho_B^x \right)^{\otimes n}. \label{eq:iid-state}
\end{equation}
A code $\cC = (\cE,\cD)$ for $\rho_{XB}^{\otimes n}$  of rate $R$ is called an $(n,R)$-code for $\rho_{XB}$. That is, we use the notation of ``$(n,R)$-code'' for the i.~i.~d. case. We are interested in the optimal (minimal) average probability of error at a fixed rate:
\begin{equation}
P_\text{e,s}^\star(n,R) \equiv P_\text{e,s}^\star(\rho_{XB},n,R) := \inf \{ \Pe(\sC) : \sC \text{ is an }(n,R)\text{-code}\}
\end{equation}
Here, the subscript~s indicates \emph{s}ource coding. The (average) probability of error $\Pe(\cC)$ for a code $\cC$ is defined in \eqref{eq:Pe_source}. These quantities are associated to the \emph{error exponents},
\begin{equation}\label{eq:source_iid_exponents}
\boxed{\begin{aligned}
e_\text{s}(n,R) &:= -\frac1n \log P_\text{e,s}^\star(n,R), & 
sc_\text{s}(n,R) &:= -\frac1n \log (1-P_\text{e,s}^\star(n,R)). 
\end{aligned}}
\end{equation}
Here  $e$ is the finite blocklength error exponent, while $sc$ is the finite blocklength strong converse exponent.
 \end{enumerate}

To describe the second form of $p(\bx)$ on $\cX^n$ that we consider, we first recall the notion of types. The set of types for sequences in $\cX^n$ is given by
\begin{align}
\mathscr{P}_n(\mathcal{X}) := \left\{  P\in\mathscr{P}(\mathcal{X}) \left| P(x) \in \left\{0,\frac1n, \frac2n,\ldots, 1 \right\},\, \forall x\in\mathcal{X}\right.
\right\}.
\end{align}
Here, $\mathscr{P}(\cX)$ is the set of probability distributions on $\cX$. A sequence $\bx \in\mathcal{X}^n$ has \emph{type} (or \emph{composition}) $Q$ if 
\begin{equation}
\frac1n \sum_{i=1}^n \mathbf{1}_{\left\{ x = x_i \right\}} = Q(x), \quad \forall x\in\mathcal{X}.
\end{equation} 
We denote by $T_{Q}^n$ the set of all such sequences in $\mathcal{X}^n$ with type $Q$.

\begin{enumerate}[(i), start=2]
 	\item \textbf{The constant type case:}
The other form for $p(\bx)$ in \eqref{eq:rho_XB-cq} that we consider is that of a uniform distribution over a type class. For some type $Q$, we consider the probability distribution on $\cX^n$ given by
\begin{equation}
p(\bx) = \begin{cases}
\frac{1}{|T_Q^n|} & \text{ if }\bx \in T_Q^n \\
0 & \text{else.}
\end{cases}
\end{equation}
In this case, \eqref{eq:rho_XB-cq} becomes
\begin{equation} \label{eq:source-of-type-Q}
\rho_{X^nB^n} = \rho_{X^nB^n}^{(Q)} := \frac{1}{|T_Q^n|} \sum_{\bx\in T_Q^n} \ket{\bx}\bra{\bx}\otimes \rho_{B^n}^{\bx}.
\end{equation}
In this case, we define an $(n,R,Q)$-code for $\rho_{XB}$ as a code $\cC = (\cE,\cD)$ of rate $R$ for $\rho_{X^nB^n}^{(Q)}$. As before, we consider the optimal probability of error over such codes, namely
\begin{equation}
P_\text{e,s}^\star(n,R,Q) \equiv P_\text{e,s}^\star(\rho_{X^nB^n}^{(Q)},n,R,Q) := \inf \{ \Pe(\sC) : \sC \text{ is an }(n,R,Q)\text{-code}\}.
\end{equation}
As before, we define the associated error exponents,
\begin{equation} \label{eq:source_type_exponents}
\boxed{\begin{aligned}
e_\text{s}(n,R,Q) &:= -\frac1n \log P_\text{e,s}^\star(n,R,Q), &   \\
sc_\text{s}(n,R, Q) &:= -\frac1n \log (1-P_\text{e,s}^\star(n,R,Q)).
\end{aligned}}
\end{equation}
\begin{remark}
The c-q state $\rho_{X^nB^n}^{(Q)}$ can be seen as the Choi-Jamio\l{}kowski state of the c-q channel $\mathscr{W}^{\otimes n}$ defined on the set of sequences of type $Q$. In contrast, in general $\rho_{XB}^{\otimes n}$ is not the Choi-Jamio\l{}kowski state of any c-q channel. This is evident from the fact that the Choi-Jamio\l{}kowski state associated to any c-q channel is a c-q state with a uniform distribution.
\end{remark}
\end{enumerate}

The states $\rho_{XB}^{\otimes n}$ and $\rho_{X^nB^n}^{(Q)}$ are related by the \emph{type decomposition},
\begin{equation} \label{eq:type-decomposition}
\rho_{XB}^{\otimes n} = \sum_{Q\in \mathscr{P}_n(\cX)} \Pr[\bx \in T_Q^n] \, \rho_{X^nB^n}^{(Q)},
\end{equation}
where $\Pr[\bx \in T_Q^n] := \sum_{\bx\in T_Q^n} p(\bx)$, and $p(\bx) = p(x_1)p(x_2)\dotsm p(x_n)$ is the i.i.d.~probability distribution from \eqref{eq:iid-state}.

It will be useful to collect some elementary results from the theory of types. 
It is well-known that \cite{CK11}
\begin{align}
|\mathscr{P}_n(\mathcal{X})| \leq (n+1)^{|\mathcal{X}},\quad \forall n\in\mathbb{N}. \label{eq:size_P}
\end{align}
Additionally, given an i.i.d.~source $X$ with distribution $P \in \mathscr{P}(\cX)$, the probability of $X$ emitting a sequence $\bx$ with type $Q$ can be bounded as \cite{CK11}
\begin{align}
(n+1)^{-|\mathcal{X}|} \exp\{ -n D(Q\|P) \}
\leq \Pr\left[ \bx \in T_{Q}^n  \right]
\leq \exp\{ -n D(Q\|P) \}. \label{eq:prob_T}
\end{align}
Further, the size of $T_{Q}^n$ satisfies the bounds \cite{CK11}
\begin{align}
(n+1)^{-|\mathcal{X}|} \exp\left\{ nH(Q)  \right\} \leq |T_{Q}^n| \leq \exp\left\{ nH(Q)  \right\}. \label{eq:size_T}
\end{align}

\subsection{Classical-Quantum Channel Coding}
A classical-quantum (c-q) channel is a map $\mathscr{W}$ from a set $\cX$ to the set $S(B)$ of density matrices on a Hilbert space $\cH_B$. An \emph{$n$-blocklength code} $\cC = (\cE,\cD)$ for a c-q channel consists of an encoder $\cE$, which is a map  from a \emph{message set} $\cM$ to $\cX^n$, and a decoder $\cD$, which is a POVM $\{\Lambda_{\hat m}\}_{\hat m\in \cM}$ on $S(B^n)$ with outcomes in $\cM$. 

The channel $\mathscr{W}$ is \emph{classical} when it has the form
\begin{equation} \label{eq:classical_channel}
\mathscr{W} : x \mapsto  \sum_{y}p(y|x)\ket{y}\bra{y}
\end{equation}
where $\{\ket{y}\}_y$ is an orthonormal basis of $\cH_B$ and $\{p(y|x)\}_y$ is a (conditional) probability distribution. In this case, $\mathscr{W}$ is equivalently described by $\{p(y|x)\}_{x,y}$.

An $n$-blocklength code defines a procedure to pass messages $m\in \cM$ through the c-q channel. The encoder maps $m$ to a \emph{codeword} $\cE(m) = (x_1,\dotsc,x_n) \in\cX^n$. The set of all such codewords, $\{\cE(m) : m\in \cM\}$, is referred to as the \emph{codebook}. The c-q channel $\mathscr{W}$ is used $n$ times, yielding the quantum state
\[
W^{\otimes n}(\bx) = W(x_1)\otimes \dotsc \otimes W(x_n)\in S(B^n).
\]
This $n$-partite state is measured using the POVM given by the decoder, which gives the outcome $\hat m$ with probability
\[
\tr[\Lambda_{\hat m} W(x_1)\otimes \dotsc \otimes W(x_n)].
\]
The probability $\Pe(\cC,m)$ of an error on message $m$ is the probability that $\hat m\neq m$, which can be written
\begin{equation}
\Pe(\cC,m) = 1 - \tr[\Lambda_{m} W^{\otimes n}( \cE(m)) ].
\end{equation}
This yields quantities describing the probability of error for a code $\cC$: the average probability of error,
\begin{equation} \label{eq:def_avg_error_chan}
\Pe(\cC) \equiv \Peavg(\cC) := \frac{1}{|\cM|} \sum_{m\in \cM} \Pe(\cC,m)
\end{equation}
and the maximal probability of error,
\begin{equation} \label{eq:def_max_error_chan}
\Pemax(\cC) := \max_{m\in \cM} \Pe(\cC,m).
\end{equation}

The \emph{rate} of the code is the number of bits of message which can be sent per use of the channel:
\begin{equation}
\rate(\cC) = \frac{\log(|\cM|)}{n}.
\end{equation}
An $(n,R)$-channel code is a $n$-blocklength code with rate $R$.

A standard technique to construct codes with small maximal error given a code with small average error called \emph{expurgation} (see, e.g. \cite[Eq. (4.41)]{SGB67}) yields the following lemma.
\begin{lemm}[Expurgation] \label{lem:expurgation}
Let $\cC$ be an $n$-blocklength c-q channel code with message set $\cM$ of size $M := |\cM|$ and average probability of error $\Peavg(\cC)$. Then there exists an $n$-blocklength c-q channel code $\tilde \cC$ with message set $\tilde \cM$ of size $\floor{\frac{M}{2}}$ and maximal probability of error
\[
\Pemax(\tilde \cC) \leq 2 \Peavg(\cC).
\]
\end{lemm}

There is a particular class of encoding maps for c-q channels which will play a distinguished role in this work. A \emph{constant composition} code of type $Q$ is an $n$-blocklength c-q channel code $\cC = (\cE,\cD)$ with the property that each codeword is of type $Q$: $\cE(m)\in T_Q^n$ for each $m\in \cM$. We note the expurgation lemma applied to a constant composition code $\cC$ of type $Q$ yields a code $\tilde \cC$ which is still constant composition and of type $Q$.

The optimal (average) error probability over constant composition codes of type $Q$ is given by
\begin{equation}
P_\text{e,c}^\star(n,R,Q) := \inf \{ \Pe(\sC) : \sC \text{ is an }(n,R)\text{-channel constant composition code of type $Q$}\}.
\end{equation}
Similarly, the finite blocklength error exponent and strong converse exponent are 
\begin{equation} \label{eq:chan_type_exponents}
\boxed{\begin{aligned}
e_\text{c}(n,R,Q) &:= -\frac1n \log P_\text{e,c}^\star(n,R,Q),  &
sc_\text{c}(n,R,Q) &:= -\frac1n \log (1-P_\text{e,c}^\star(n,R,Q)).  
\end{aligned}}
\end{equation}

\section{Operational Duality} \label{sec:operational_duality}
First, let us write the classical operational duality result of Ahlswede and Dueck given by Theorem~1 of \cite{AD82} in the ``quantum notation'' of  Section~\ref{sec:prelim}.
\begin{theo*}[\textbf{Ahlswede and Dueck}, \cite{AD82}] \label{thm:classical-duality}
An i.i.d. classical source with classical side information is described the joint state
\[
\rho_{XB} = \sum_{x\in \cX}p(x) \ket{x}\bra{x} \otimes \rho_B^x
\]
where $\rho_B^x$ satisfies \eqref{eq:CSI}.
 Then the map $\mathscr{W}$ defined by
\[
\mathscr{W}: x\mapsto \rho_B^x = \sum_y p(y|x)\, \ket{y}\bra{y}
\]
is a classical channel (i.e.~satisfies \eqref{eq:classical_channel}). The optimal error exponent $e_\mathnormal{s}(n,R)$ for the  source coding of $\rho_{XB}$ at a rate $R$, and the optimal error exponent $e_\textnormal{c}(n, R', Q)$ for the channel coding of $\mathscr{W}$ at a rate $R'$ (using constant composition codes of type $Q$), can be related as follows.
For any $\delta>0$ there exists $N_0 \equiv N_0(\delta, |\cX|, \dim \cH_B)\in \mathbb{N}$ such that for all $n\geq N_0$, 
\begin{align}
 e_\textnormal{s}(n, R+\delta) &\geq \min_{Q\in \mathscr{P}_n} [ D(Q\|P)+ e_\textnormal{c}(n,H(P)-R, Q)] - \delta\\
 e_\textnormal{s}(n, R) &\leq \min_{Q\in \mathscr{P}_n} [ D(Q\|P)+ e_\textnormal{c}(n,H(P)-R - \delta, Q)] + \delta
\end{align}
where $P$ denotes the distribution $\{p(x)\}_{x\in \cX}$. 
\end{theo*}

The following result is a generalization of the above theorem to the case when the classical source has quantum side information (i.e.~\eqref{eq:CSI} need not hold), and consequently the induced channel $\mathscr{W} : x \mapsto \rho_B^x$ is a (general) c-q channel, along with an explicit determination of the dependence of $\delta$ on $n$.
\begin{theo}[Operational duality of the error exponents] \label{thm:e-duality}
An i.i.d.~classical source with quantum side information is described by the joint state
\begin{equation}
\rho_{XB} = \sum_{x\in \cX} p(x) \ket{x}\bra{x}\otimes \rho_B^x
\end{equation}
where $\rho_B^x \in S(B)$. Then the map from $\cX\to S(B)$ defined by $\mathscr{W} : x \mapsto \rho_B^x$ is a c-q channel. The optimal error exponent $e_\mathnormal{s}(n,R)$ for the  source coding of $\rho_{XB}$ at a rate $R$, and the optimal error exponent $e_\textnormal{c}(n,R', Q)$ for the channel coding of $\mathscr{W}$ at a rate $R'$ (using constant composition codes of type $Q$), can be related as follows. For any $n\in \mathbb{N}$,
\begin{align}
	e_\textnormal{s}(n,R+ \delta_n' + \delta_n) &\geq \min_{Q\in\mathscr{P}_n(\mathcal{X})} \left[ D(Q\|P) + e_\textnormal{c}(n, H(Q)-R, Q) \right]  -  \delta_n'. \label{eq:duality_ac2}\\
	e\source(n,R) &\leq \min_{Q\in\mathscr{P}_n(\mathcal{X})} \left[ D(Q\|P) + e_\textnormal{c}(n, H(Q)-R - \delta_n', Q) \right]  +\delta_n' . \label{eq:duality_op2}
\end{align}
	where $P$ denotes the distribution $\{p(x)\}_{x\in \cX}$, and
	\begin{equation} \label{eq:def_delta_deltaprime}
	\delta_n := \frac{1}{n}\log \left( 2n \log(|\cX|) + 1\right), \qquad \delta_n' :=(|\cX|+1) \frac{\log(n+1)}{n}.
	\end{equation}
\end{theo}

The above theorem builds on a type-dependent duality given by the following result.
\begin{theo}[Operational duality of the error exponents at fixed type $Q$] \label{thm:e-duality-constant-type}
For any $n\in \mathbb{N}$, and type $Q\in\mathscr{P}_n(\mathcal{X})$, and $R \geq \delta_n$,
	\begin{align} \label{eq:ec-es-type-duality}
	e_\textnormal{c}(n, H(Q)-R+\delta_n, Q)  - \frac{1}{n} \leq e_\textnormal{s}(n,R,Q) \leq  e_\textnormal{c}(n,H(Q)-R-\delta_n', Q)  +\frac{1}{n}\log(1+\frac{1}{n})  
	\end{align}
	where $\delta_n$ and $\delta_n'$ are defined in \eqref{eq:def_delta_deltaprime}.
\end{theo}
\begin{remark}
If the families of functions $\{R\mapsto e_\textnormal{c}(n, R, Q)\}_{n \in \mathbb{N}}$ and $\{R\mapsto e_\textnormal{s}(n, R, Q)\}_{n \in \mathbb{N}}$  are each equicontinuous, then Theorem~\ref{thm:e-duality-constant-type} implies the asymptotic error exponents satisfy
\begin{equation} \label{eq:asymp-e-type-duality}
e_\textnormal{s}(R,Q) = e_\textnormal{c}(H(Q)-R, Q).
\end{equation}
\end{remark}
We prove the lower bounds on $e\source(n,R)$ and $e\source(n,R,Q)$ in Theorems~\ref{thm:e-duality} and \ref{thm:e-duality-constant-type}  in Section~\ref{sec:source-from-channel}, and the corresponding upper bounds in Section~\ref{sec:chan-from-source}.

\subsection{Building source codes from channel codes\label{sec:source-from-channel}}

We will establish the lower bound on $e_\textnormal{s}(n,R)$ given in \eqref{eq:duality_ac2}  and  the lower bound on $e_\textnormal{s}(n,R,Q)$ given in \eqref{eq:ec-es-type-duality} by constructing CQSW source codes from c-q channel codes for the side information. The main tool for building source codes from channel codes will be the following type covering lemma, which is a simple strengthening of the one due to Ahlswede \cite[Section 5.6.1]{Ahl80}.

Before starting the proof, let us first outline the construction in the constant-type case. We are given a constant-composition channel code, which has a message set $\cM$, an encoder $\cE\chan$, a resulting codebook $\cE\chan(\cM) \subset T_Q^n$, and a decoder, which aims to discriminate between the states $\{\rho_B^\bx : \bx \in \cE\chan(\cM)\}$. From this, we aim to construct a source code, which has an encoder $\cE\source: T_Q^n \to \cZ$ for some compressed set $\cZ$, and a decoder, which given $z\in \cZ$ aims to discriminate between the states in the set $S_z :=\{ \rho_B^x : \bx\in \cE\source\inv(\{z\})\}$. The set $S_z$ constitutes the possibilities for the QSI when the message is compressed as $z$.
The channel decoder only knows how to discriminate between states in one subset, i.e.\ those corresponding to elements of the codebook, while the source decoder has to discriminate between states in many such sets, namely between states in $S_z$  for each compressed message $z\in \cZ$, and moreover these sets must cover all of $\{\rho_B^\bx:\bx\in T_Q^n\}$, a strict superset of the codebook. So how does one define the source decoder from the more limited channel decoder? Recall that all elements of $T_Q^n$ can be related to each other by permutations. We will essentially choose the sets $S_z$ to be a permutation $\pi_z$ of the codebook. Since the source decoder knows the value of $z$, it can invert the permutation and apply the channel decoder to the resulting element of the codebook. Thus, the number of permutations of the codebook required to cover all of $\{\rho_B^\bx:\bx\in T_Q^n\}$ determines the size of the compressed space $\cZ$. We will first discuss the type covering lemma used to bound the number of required permutations, then proceed to the formal proof.

Given a type $Q$ and a subset $\mathcal{U}\subset T_Q^n$, we are interested in sets of permutations of the numbers $\{1,\dotsc,n\}$ which induce a \emph{cover} of $T_Q^n$: that is, sets of permutations $\{\pi_1,\dotsc, \pi_L\}$ for $L\in \mathbb{N}$, such that
\[
\bigcup_{i=1}^{L} \pi_{i} \,\mathcal{U} = T_{Q}^n,
\] 
using the notation $\pi_{i} \mathcal{U} := \left\{ \pi_i \mathbf{u}: \mathbf{u}\in \mathcal{U} \right\}$.
We note if $|\mathcal{U}|=1$, then $L = |T_Q^n|$ is necessary and sufficient. The following type covering lemma provides a more useful bound on $L$ when $|\mathcal{U}|>1$.
\begin{lemm}[Type covering lemma] \label{lem:covering}
	For any type $Q$, integer $n$, and nonempty subset $\mathcal{U}\subset T_Q^n$, there exists a set of $L_Q := \ceil{ |\mathcal{U}|^{-1} |T_Q^n| \log |T_Q^n|  }$ permutations $\pi_1, \ldots, \pi_{L_Q}$ of the integers $\{1,\ldots, n\}$ which induce a cover of $T_Q^n$:
	\begin{align}
	\bigcup_{i=1}^{L_Q} \pi_{i}\, \mathcal{U} = T_{Q}^n.
	\end{align}
	
Moreover, if one draws $2L_Q$ permutations  independently and uniformly at random from the set of all permutations of $\{1,\dotsc,n\}$, the resulting collection $\{\pi_1,\ldots,\pi_{2L_Q}\}$ will induce a cover $T_Q^n$ with probability at least $1 - \frac{1}{|T_Q^n|}$.
	In particular, the expected number of draws of $2L_Q$ permutations required to find a cover of $T_Q^n$ is less than $2$ for any type $Q$, and any $n\geq 2$.
	\end{lemm}
	\begin{remark}This lemma provides a more constructive version of the original covering lemma by Ahlswede (which is an existence result). Note, in the sequel, we will be interested in rates of the form $\frac{1}{n}\log(L_Q)$, which remain unaffected by the replacement $L_Q\to 2L_Q$ in the asymptotic limit $n\to\infty$. The proof is a simple adaptation of Ahlswede's proof, and is included in Appendix~\ref{sec:proof-covering} for completeness.
	\end{remark}

With this result in hand, we proceed to first prove the lower bound of Eq.~\eqref{eq:ec-es-type-duality} for a fixed type-dependent source, and then Eq.~\eqref{eq:duality_ac2} for an i.i.d.~source. This argument follows the  of the proof of Theorem~1 of Ahlswede and Dueck~\cite{AD82}, and Eq.~\eqref{eq:perm-as-sym} below plays the main role in employing the type covering lemma in the quantum case.
	
Fix $R>0$, $n \in \mathbb{N}$, and an arbitrary type $Q \in \mathscr{P}_n(\mathcal{X})$. We will consider an $n$-blocklength c-q channel code $\cC\chan = (\cE\chan,\cD\chan)$ which is constant composition of type $Q$, and has rate $H(Q)-R$. By the expurgation method (Lemma~\ref{lem:expurgation}), there is a message set $\cM$ with size
	\begin{align}
	M := |\cM| \geq \frac12 \exp\{ n (H(Q)-R) \}, \label{eq:duality_ac5}
	\end{align}
	 such that the codewords $\mathcal{U} := \cE\chan(\cM) = \{\mathbf{u}_1, \ldots, \mathbf{u}_M \} \subset T_{Q}^n$ and a decoder, i.e.~POVM, $(\Lambda_{m})_{m\in \cM}$,  such that the maximal error probability for these codewords can be upper bounded by the average error probability of $\cC\chan$:
	\begin{equation}
	\max_{m \in \cM} \left(1 - \Tr\left[W^{\otimes n}(\cE\chan(m)) \Lambda_{m} \right]\right) \leq 2 \Pe(\cC\chan). \label{eq:chan-max-error-bound} 
	\end{equation}
	From this channel code, we construct a Slepian-Wolf code $\cC\source = (\mathcal{E}\source, \mathcal{D}\source)$, for the $n$-blocklength source $\rho^{(n)}_{X^nB^n}$  associated to the constant composition ensemble of type $Q$ (see \eqref{eq:source-of-type-Q}), as follows.

We apply Ahlswede's covering lemma, Lemma~\ref{lem:covering}, with $\mathcal{U}:=\cE\chan(\cM) \subset T_Q^n$, yielding a set of permutations $\{\pi_1,\dotsc,\pi_{L_Q}\}$ for
	\begin{equation}
	\bigcup_{i=1}^{L_Q} \pi_i \mathcal{U} = T_Q^n
	, \qquad L_Q = \ceil{ M^{-1} |T_Q^n| \log |T_Q^n|  }. \label{eq:duality_ac6}
	\end{equation}
	Note that every permutation $\pi_i$ will induce a unique unitary $V_i$ on $\mathcal{S}(B^{ n})$ which permutes the Hilbert spaces, i.e.~
	\begin{equation} \label{eq:perm-as-sym}
	W^{\otimes n}(\pi_i \mathbf{u}) = V_i W^{\otimes n}(\mathbf{u}) V_i^\dagger, \qquad \forall \, \mathbf{u} \in \mathcal{U}, \; \forall \, i \in [L_Q].
	\end{equation}
	We partition the set $|T_Q^n|$ by defining $S_{1,Q}= \pi_1 \mathcal{U}$, and sequentially defining
	\begin{equation}
	S_{i,Q} := \pi_i \mathcal{U} - \bigcup_{ \iota =1}^{i-1} \pi_{ \iota } \mathcal{U}. \label{eq:duality_ac7}
	\end{equation}
	We use this partition to define the source encoder $\mathcal{E}\source$ by
	\begin{equation}
	\begin{aligned}
	\cE\source : \quad  T_Q^n &\to \cZ := [L_Q]\\
	 \bx &\mapsto i, \quad \text{if } \bx \in S_{i,Q},
	\end{aligned}
	\end{equation}
	where $[L_Q] = \{1,\dotsc,L_Q\}$. We define the source decoder by $\cD\source = \{ \cD\source^{(i)} \}_{i \in [L_Q]}$ where $\cD\source^{(i)} $ is a POVM defined as follows. For each $i\in [L_Q]$ and $\bx\in T_Q^n$, we set
	\begin{equation}
	\Pi_{\bx }^{(i)} := \begin{cases}
	V_i\Lambda_{m} V_i^\dagger & \text{if }  \bx = \pi_i \cE\chan(m) \in S_{i,Q}\\
	0 & \text{if } \bx \not \in S_{i,Q}.
	\end{cases}
	\end{equation}
	Note that
	\begin{equation}
	\sum_{\bx\in T_Q^n}\Pi_{\bx }^{(i)} = \sum_{m : \pi_i \cE\chan(m)\in S_{i,Q}} V_i\Lambda_{m} V_i^\dagger \leq \sum_{m\in \cM}V_i\Lambda_{m} V_i^\dagger 
	= V_i \one V_i^\dagger 
	= \one
	\end{equation}
	using that $S_{i,Q} \subset \pi_i \mathcal{U}$ in the inequality. Then we can recover a true POVM $\{\tilde \Pi^{(i)}_\bx\}_{\bx \in T_Q^n}$, i.e.~ such that $\sum_{ \bx \in T_Q^n } \tilde\Pi_{\bx }^{(i)} = \one$, by choosing any $\bx_0\in \cX^n$ and defining
	\[
	\tilde \Pi^{(i)}_{\bx} = \begin{cases}
	\Pi^{(i)}_{\bx} & \bx \neq \bx_0\\
	\one - \sum_{ \bx \in S_{i,Q} } \Pi_{\bx }^{(i)} & \bx = \bx_0.
	\end{cases}
	\]
We therefore define $\cD\source^{(i)} = \{\tilde \Pi^{(i)}_\bx\}_{\bx \in \cX^n}$. The operator inequality $\tilde \Pi^{(i)}_\bx \geq \Pi^{(i)}_\bx$ for $\bx \in T_Q^n$ shows that the probability of error under $\{\tilde \Pi^{(i)}_\bx\}_{\bx \in T_Q^n}$ is not greater than the probability of error under $\{\Pi^{(i)}\}_{\bx \in T_Q^n}$, so we may simply consider decoding with the subnormalized set $\{\Pi^{(i)}_\bx\}_{\bx \in T_Q^n}$.
	
Now, we compute the rate and the error probability of the code $\cC\source$.	Equations \eqref{eq:duality_ac5} and \eqref{eq:size_T} imply that
	\begin{align}
	|\mathcal{E}\source(\cX^n)| &= L_Q \\
	&\leq 2 \exp\{ -n (H(Q)-R) + n H(Q)  \}\cdot nH(Q) + 1 \\
	&\leq \exp\{ n(R+\delta_n)\} \label{eq:source-rate-bound}
	\end{align}
	using $H(Q)\leq \log |\cX|$ and $\e^{-nR}\leq 1$, and setting $	\delta_n = \frac{1}{n}\log \left( 2n \log |\cX| +1 \right)$.
	Next, we bound the probability of error of the source code by relating it to the maximal probability of error of the channel code, and using the expurgated error bound \eqref{eq:chan-max-error-bound}. By definition,
	\begin{align}
	\Pe(\mathcal{C}\source) &= 1 - \frac{1}{|T_Q^n|} \sum_{\bx\in T_Q^n} \tr[ \rho_\bx^{B^n} \Pi^{(\cE\source(\bx))}_{\bx} ]\\
	&=1 - \frac{1}{|T_Q^n|} \sum_{i=1}^{L_Q} \sum_{\bx\in S_{i,Q}} \tr[ \rho_\bx^{B^n} \Pi^{(\cE\source(\bx))}_{\bx} ]
	\end{align}
	using that $\bigsqcup_{i=1}^{L_Q} S_{i,Q} = T_Q^n$. If $\bx \in S_{i,Q}$ then $\bx = \pi_i\cE\chan(m)$ for exactly one message $m\in \cM$, which we will denote as $m_\bx$. Then $\rho_{\bx}^{B^n} = W^{\otimes n}(\pi_i\cE\chan (m_\bx)) = V_i W^{\otimes n}(\cE\chan(m_\bx))V_i^\dagger$, since $W^{\otimes n} : \cX^n \to S(B^n)$ maps $\bx$ to the side information associated to $\bx$, namely $\rho_{\bx}^{B^n}$. Additionally, $\bx\in S_{i,Q}$ is equivalent to $\cE\source(\bx)=i$, so $\Pi^{(\cE\source(\bx))}_{\bx} = V_i \Lambda_{m_\bx}V_i^\dagger$, and 
	\begin{align}
	\Pe(\mathcal{C}\source) &=1 - \frac{1}{|T_Q^n|} \sum_{i=1}^{L_Q} \sum_{\bx\in S_{i,Q}} \tr[V_i W^{\otimes n}(\cE\chan(m_\bx))V_i^\dagger V_i \Lambda_{m_\bx}V_i^\dagger] \\
	&=1 - \frac{1}{|T_Q^n|} \sum_{i=1}^{L_Q} \sum_{\bx\in S_{i,Q}} \tr[W^{\otimes n}(\cE\chan(m_\bx))  \Lambda_{m_\bx}] 
	\end{align}
by cyclicity of the trace and that $V_i^\dagger V_i = V_i^\dagger V_i = \one$. Then, bounding the error of the message $m_\bx$ by the maximal error,
\begin{align}
\Pe(\mathcal{C}\source) &=\frac{1}{|T_Q^n|} \sum_{i=1}^{L_Q} \sum_{\bx\in S_{i,Q}} \left( 1-\tr[W^{\otimes n}(\cE\chan(m_\bx))  \Lambda_{m_\bx}]  \right)\\
&\leq \frac{1}{|T_Q^n|} \sum_{i=1}^{L_Q} \sum_{\bx\in S_{i,Q}} \max_{m\in \cM} \left( 1-\tr[W^{\otimes n}(\cE\chan(m))  \Lambda_{m}]  \right)\\
&= \max_{m\in \cM} \left( 1-\tr[W^{\otimes n}(\cE\chan(m))  \Lambda_{m}]  \right) \label{eq:collapse-sum}\\
&\leq 2 \Pe(\cC\chan) \label{eq:bound-by-chan}
\end{align}
using in \eqref{eq:collapse-sum} that $\{ S_{i,Q}\}_{i=1}^{L_Q}$ is a partition of $T_Q^n$ and using the expurgated bound \eqref{eq:chan-max-error-bound} in \eqref{eq:bound-by-chan}. Thus, since the optimal source code probability of error is at most the the probability of error of this code, we have
\begin{equation} \label{eq:Pestarsource-bound-chan}
\Pestarsource(n,R + \delta_n,Q) \leq 2 \Pe(\cC\chan) .
\end{equation}
 Since this holds for any channel code of rate $H(Q)-R$ and of constant composition $Q$, we may minimize over $\cC\chan$ to find
\begin{equation} \label{eq:source-error-UB}
\Pestarsource(n,R + \delta_n,Q)  \leq 2 \Pestarchan(n, H(Q)-R, Q).
\end{equation}
This yields the lower bound Eq.~\eqref{eq:ec-es-type-duality}:
	\begin{align}
	e_\text{s}(n,R+\delta_n,Q) \geq  e_\textnormal{c}(n, H(Q)-R, Q)  - \frac{1}{n}\log 2.
	\end{align}
	
Next, we move on to prove the second claim, Eq.~\eqref{eq:duality_ac2}. The task is to construct a source code for an i.i.d.~CQSW source $\rho_{XB}^{\otimes n}$ with distribution $P$. For each type $Q\in \mathscr{P}_n$, we  consider an arbitrary channel code $\cC\chan^{(Q)} = (\cE\chan^{(Q)}, \cD\chan^{(Q)})$ of constant composition $Q$, with rate $H(Q)-R$ and message set $\cM^{(Q)}$. Then we construct a corresponding source code $\cC\source^{(Q)} = (\cE\source^{(Q)}, \cD\source^{(Q)})$ as before, with
	\begin{equation} \label{eq:Q-bound-s-by-c}
	\Pe(\cC\source^{(Q)} ) \leq 2 \Pe(\cC\chan^{(Q)})
	\end{equation}
by Eq.~\eqref{eq:bound-by-chan}. We define the i.i.d.~source encoder as
	\begin{equation}
	\begin{aligned}
	\cE\source : \quad \cX^n &\to  \cZ := \left\{ (i,Q) :i\in [L_Q], Q\in \mathscr{P}_n(\cX) \right\}\\
	 \bx &\mapsto  (\cE\source^{(P_\bx)}(\bx),P_\bx)
	\end{aligned}
	\end{equation}
where $P_\bx$ denotes the type of the sequence $\bx$. The type-dependent source decoder $ \cD\source^{(Q)} $ is given by a collection of POVMs, $ \cD\source^{(Q)} = \{ \cD^{(i,Q)} \}_{i\in [L_Q]}$, and we define the type-independent source decoder by
\begin{equation}
\cD\source = \bigcup_{Q\in \mathscr{P}_n} \cD\source^{(Q)} = \{\cD^{(i,Q)} : i\in [L_Q], Q\in\mathscr{P}_n(\cX) \}, \qquad \cD^{(i,Q)} = \{\Pi_\bx^{(i,Q)}\}_{\bx\in\cX^n}.
\end{equation}
Then we have the cardinality bound
 \begin{align}
	|\cE\source(\cX^n)| &\leq | \mathscr{P}_n(\mathcal{X})  |\max_{Q\in\mathscr{P}_n(\mathcal{X}) } L_Q \\
	&\leq (n+1)^{|\mathcal{X}|} \max_{Q\in\mathscr{P}_n(\mathcal{X}) } \left[ 2  \exp\{ -n (H(Q)-R) + n H(Q)  \}\cdot nH(Q) + 1 \right] \\
	&\leq \exp\{ n(R+\delta_n + \delta_n')\}
\end{align}
for $\delta_n' = \frac{|\cX|+1}{n}\log(n+1),$ where we recall Eq.~\eqref{eq:size_P} in the first inequality. This yields $\rate(\cC\source) \leq R + \delta_n +  \delta_n'$. Additionally,
\begin{align}
\Pe(\cC\source) &=1- \sum_{\bx\in \cX^n}p(\bx) \tr[\Pi_{\bx}^{(\cE\source(\bx))}\rho_\bx^{B^n}]\\
&=1- \sum_{Q\in \mathscr{P}_n(\cX^n)} \sum_{\bx\in T_Q^n} p(\bx) \tr[\Pi_{\bx}^{(\cE\source^{(Q)}(\bx), Q)}\rho_\bx^{B^n}].
\end{align}
Any sequence $\bx$ of type $Q$ has the same probability $p(\bx)$, since the source is i.i.d.. Then writing $\Pr(\by\in T_Q^n) = \sum_{\by\in T_Q^n} p(\by) = |T_Q^n| p(\bx)$, we have
\begin{equation} \label{eq:pr-Q}
p(\bx)= \frac{1}{|T_Q^n|}\Pr(\by\in T_Q^n), \qquad \forall \,\bx \in T_Q^n.
\end{equation}
Thus,
\begin{align}
\Pe(\cC\source)&=1- \sum_{Q\in \mathscr{P}_n(\cX^n)}\Pr(\bx\in T_Q^n) \frac{1}{|T_Q^n|}\sum_{\bx\in T_Q^n} \tr[\Pi_{\bx}^{(\cE\source^{(Q)}(\bx), Q)}\rho_\bx^{B^n}]\\
&=\sum_{Q\in \mathscr{P}_n(\cX^n)}\Pr(\bx\in T_Q^n) \Pe(\cC\source^{(Q)})\\
&\leq 2\sum_{Q\in \mathscr{P}_n(\cX^n)}\Pr(\bx\in T_Q^n) \Pe(\cC\chan^{(Q)})\\
&\leq 2\sum_{Q\in \mathscr{P}_n(\cX^n)}\Pr(\bx\in T_Q^n) \Pestarchan(n, H(Q)-R,Q)
\end{align}
using \eqref{eq:Q-bound-s-by-c} and minimizing over the channel codes $\{\cC\chan^{(Q)}\}_{Q\in \mathscr{P}_n(\cX^n)}$. Then
\begin{equation}
\Pe(\cC\source) \leq  2 |\mathscr{P}_n(\cX^n)| \; \max_{Q\in \mathscr{P}_n(\cX^n)} \Pr(\bx\in T_Q^n) \, \Pestarchan(n, H(Q)-R,Q).
\end{equation}
Since the optimal source code has probability of error at most that of $\cC\source$, we have
\begin{align}
\Pestarsource(n,R + \delta_n + \delta_n') &\leq  2|\mathscr{P}_n(\cX^n)| \; \max_{Q\in \mathscr{P}_n(\cX^n)} \Pr(\bx\in T_Q^n) \, \Pestarchan(n, H(Q)-R,Q)\\
&\leq 2|\mathscr{P}_n(\cX^n)| \max_{Q\in \mathscr{P}_n(\cX^n)} \exp \Big(-n (D(Q\|P) + e\chan(n,H(Q)-R, Q)) \Big)
\end{align}
using \eqref{eq:prob_T} and the definition of $e\chan(n,H(Q)-R, Q)$. Then by the definition of $e\source(n,R)$,
\begin{align}
e\source(n,R + \delta_n + \delta_n') &\geq - \frac{1}{n}\log(2) - \frac{1}{n} \log |\mathscr{P}_n(\cX^n)|  + \min_{Q\in \mathscr{P}_n(\cX^n)}[ D(Q\|P) + e\chan(n,H(Q)-R, Q)]\\
&\geq - \frac{|\cX| +1}{n}\log(n+1) + \min_{Q\in \mathscr{P}_n(\cX^n)}[D(Q\|P) + e\chan(n,H(Q)-R, Q)]
\end{align}
using \eqref{eq:size_P}. This proves Eq.~\eqref{eq:duality_ac2}.

\subsection{Building channel codes from source codes} \label{sec:chan-from-source} We will first prove the upper bound of \eqref{eq:ec-es-type-duality}, then use the constant type codes we develop in that proof to establish \eqref{eq:duality_op2}. 
Let us first outline the construction in the constant-type case. We are given a source code $\cC\source$ consisting of an encoder $\cE\source: T_Q^n\to \cZ$, and decoder, which, given $z\in \cZ$, discriminates between $\{\rho_B^\bx: \bx\in \cE\source\inv(\{z\})\}$. Given a message set $\cM$, we aim to construct a channel code $\cC\chan$ consisting of an encoder $\cE\chan: \cM\to T_Q^n$, and decoder $\cD\chan$ which must discriminate between states in the set $\{\rho_B^\bx: \bx\in \cE\chan(\cM)\}$.

For any $z\in \cZ$, the source decoder can discriminate between the states of $\{\rho_B^\bx: \bx\in \cE\source\inv(\{z\})\}$. Thus, each choice of $z\in \cZ$ yields a candidate code $\cC\chan^{(z)}$, with codebook $\cE\source\inv(\{z\})$, and decoder given by the source decoder conditioned on $z$. But which $z$ do we choose? There are two main constraints:
\begin{itemize}
	\item We aim to bound $\Pe(\cC\chan^{(z)})$ in terms of $\Pe(\cC\source)$, which is the probability of error averaged over the source distribution. If a particular $z$ is unlikely, according to the source distribution and source encoder, then $\Pe(\cC\chan^{(z)})$ could be much higher than $\Pe(\cC\source)$. So we need to choose an element $z$ such that $\Pe(\cC\chan^{(z)})$ can be suitably bounded.
	\item The rate of the code $\cC\chan^{(z)}$ is determined by the size of the message set, which in turn is governed by the size of the codebook. We thus need to choose $z$ with $|\cE\source\inv(\{z\})|$ large enough.
\end{itemize}
Therefore, we employ a sequence of pigeonhole arguments to find a suitable element $z\in \cZ$.

Let us begin the proof. Consider a CQSW code $\cC\source=(\cE\source,\cD\source)$ for the source of constant type $Q$, of rate $R$. The encoder is a map $\cE\source: T_Q^n\to \cZ$, with $|\cZ| = 2^{nR}$. Let $S_i = \cE\source\inv(\{i\})$ for $i\in \cZ$.  We will proceed in three steps; first, two pigeonhole arguments identify a set of good codewords, followed by the construction and analysis of the channel code.
\begin{enumerate}
	\item 
We assert that for any $m>1$, there exists an non-empty set $\widetilde{\cZ} \subseteq \mathcal{Z}$ satisfying
\begin{align}
\forall j\in \widetilde{\mathcal{Z}}, \quad |S_j| &\geq \frac1m |T_{Q}^n| \cdot |\mathcal{Z} |^{-1} ; \label{eq:cond1} \\
\sum_{j\in\widetilde{\mathcal{Z}}} |S_j| &\geq \frac{m-1}{m} |T_{Q}^n|. \label{eq:cond2}
\end{align}
To see this, let us assume there is no subset of $\mathcal{Z}$ satisfying Eq.~\eqref{eq:cond1}. Then since even singleton sets $\widetilde{\cZ} = \{j\}$ cannot satisfy Eq.~\eqref{eq:cond1}, we must have $|S_j| < \frac1m |T_{Q}^n| \cdot |\cZ |^{-1}$ for each $j \in \cZ$.
Summing over $j \in \cZ$ yields $|T_{Q}^n|=\sum_{j\in\cZ} |S_j| < \frac1m |T_{Q}^n|$, which is a contradiction. Thus, there must exist a nonempty $\widetilde{\cZ} \subseteq \cZ$ which satisfies Eq.~\eqref{eq:cond1}. Choose
\begin{equation}
 \widetilde{\cZ} = \left\{ j\in \cZ : |S_j| \geq \frac{1}{m}|T^n_{Q}| \cdot |\cZ|^{-1}\right\}
 \end{equation}
 so that $ |S_j| < \frac1m |T_{Q}^n| \cdot |\cZ |^{-1}$ for all $j \in \cZ\backslash \widetilde{\cZ}$.
This implies 
\begin{align}
\sum_{j\in  \cZ\backslash\widetilde{\cZ}} |S_j| <\frac1m |T_{Q}^n| \cdot |\cZ |^{-1}\cdot |\cZ\backslash \widetilde{W}| \leq \frac1m |T_{Q}^n|.
\end{align}
If Eq.~\eqref{eq:cond2} is violated, i.e.~$\sum_{j\in\widetilde{\cZ}} |S_j| < \frac{m-1}{m} |T_{Q}^n|$, then
\begin{align}
\sum_{j\in\cZ} |S_j| <  \frac{m-1}{m} |T_{Q}^n| + \frac{1}{m}|T_Q^n| = |T_{Q}^n|,
\end{align}
which also leads to a contradiction. 
We have thus shown the existence of an non-empty $\widetilde{\cZ} \subseteq \cZ$ satisfying Eqs.~\eqref{eq:cond1} and \eqref{eq:cond2}. This argument appears in the case $m=2$ in \cite{AD82}.

\item Next, we define
\begin{equation}
q_{\bx} = \frac{1 - \tr[ \Pi_{\bx}^{\cE(\bx)} \rho_{B^n}^{\bx} ]}{\sum_{\by\in T_Q^n}(1 - \tr[ \Pi_{\by}^{\cE(\by)} \rho_{B^n}^{\by} ])}
\end{equation}
as the fraction of the total error incurred by source output $\bx$. Note $\{q_\bx\}_{\bx\in T_Q^n}$ forms a probability distribution. Define
\begin{equation}
Q_i := \sum_{\bx \in S_i} q_\bx.
\end{equation}
We assert there exists $j\in \widetilde \cZ$ such that
\begin{equation}
Q_j \leq \frac{m}{m-1}\frac{|S_j|}{|T_Q^n|}.
\end{equation}
Otherwise, for each $i\in \widetilde\cZ$, $Q_j > \frac{m}{m-1} \frac{|S_j|}{|T_Q^n|}$. In that case,
\begin{equation}
\sum_{j\in \widetilde \cZ} Q_j > \frac{m}{m-1}\frac{1}{|T_Q^n|} \sum_{j\in \widetilde \cZ} |S_j| \geq \frac{1}{|T_Q^n|} \frac{m}{m-1} \frac{m-1}{m}|T_Q^n| = 1
\end{equation}
which is impossible since
\begin{equation}
\sum_{j\in \widetilde \cZ}Q_j \leq \sum_{j\in \cZ}Q_j = \sum_{\bx\in T_Q^n}q_\bx = 1.
\end{equation}
\item Call this particular set $S_j$ by $\cM$. Therefore, we've identified a set $\cM  \subset T_Q^n$ such that:
\begin{equation}
|\cM | \geq \frac{1}{m}\frac{|T_Q^n|}{|\cZ|}
\end{equation}
and
\begin{equation}
\sum_{\bx \in \cM } (1 - \tr[\Pi_{\bx} \rho_{B^n}^\bx])\leq \frac{m}{m-1}\frac{ |\cM|}{|T_Q^n|} \sum_{\by\in T_Q^n}(1 - \tr[ \Pi_{\by}^{\cE(\by)} \rho_{B^n}^{\by} ]) = \frac{m}{m-1} |\cM| \Pe(\cC)
\end{equation}
for $\Pi_{\bx} := \Pi_\bx^{(j)}$. We therefore define a constant-composition channel code $\cC\chan$ of type $Q$ with message set $\cM$ and decoder $\{\Pi_\bx\}_{\bx\in \cM}$. The average probability of error is
\begin{equation}
\Pe(\cC\chan) = \frac{1}{\cM}\sum_{\bx \in \cM } (1 - \tr[\Pi_{\bx} \rho_{B^n}^\bx]) \leq \frac{m}{m-1} \Pe(\cC\source)
\end{equation}
and the rate is
\begin{align}
\rate(\cC\chan) = \frac{\log(|\cM|)}{n} &\geq -\frac{\log(m)}{n} + \frac{\log(|T_Q^n|)}{n} - \frac{\log(|\cZ|)}{n}\\
&\geq H(Q) - R -\frac{\log(m)}{n}  -|\cX| \frac{\log(n+1)}{n}.
\end{align}
Taking $m=n+1$, we have
\begin{align}
\rate(\cC\chan) &\geq H(Q) - R  -(|\cX|+1) \frac{\log(n+1)}{n}, \\
\Pe(\cC\chan)&\leq \left( 1 + \frac{1}{n} \right) \Pe(\cC\source).
\end{align}
\end{enumerate}
Thus, for any source code $\cC\source$ for a source of constant type $Q$ and rate $R$, we can construct a constant-composition channel code $\cC\chan$ of type $Q$ and rate $H(Q)-R -\delta_n$, for $\delta_n =(|\cX|+1) \frac{\log(n+1)}{n}$.
Thus, we have
\begin{equation} \label{eq:bound-Pstarchan-by-type-code}
\Pestarchan(n,H(Q)-R-\delta_n, Q) \leq \Pe(\cC\chan)\leq \left( 1 + \frac{1}{n} \right) \Pe(\cC\source)
\end{equation}
since the optimal probability of error is at most $\Pe(\cC\chan)$. 
Since this is true for any code $\cC\source$, we have
\[
\Pestarchan(n,H(Q)-R-\delta_n, Q) \leq \left( 1 + \frac{1}{n} \right) \Pestarsource(n,R,Q).
\]
Thus,
\[
e\chan(n,H(Q)-R-\delta_n, Q) \geq - \frac{1}{n}\log \left( 1+\frac{1}{n} \right) + e\source(n,R,Q).
\]
This proves the upper bound of \eqref{eq:ec-es-type-duality}.

Next, consider a CQSW code $\cC\source = (\cE\source, \cD\source)$ of rate $R$ for an i.i.d. source
\begin{align}
\rho_{X^nB^n} &= (\rho_{XB})^{\otimes n} = \sum_{\bx \in \cX^n} p(\bx) \ket{\bx}\bra{\bx} \otimes \rho^{B^n}_{\bx}\\
&= \sum_{Q\in \mathscr{P}_n(\cX^n)} \Pr(\by\in T_Q^n) \frac{1}{|T_Q^n|}\sum_{\bx\in T_Q^n} \ket{\bx}\bra{\bx}\otimes  \rho_{\bx}^{B^n}\\
&=\sum_{Q\in \mathscr{P}_n(\cX^n)} \Pr(\by\in T_Q^n)\, \rho^{Q}_{X^n B^n}
\end{align}
using \eqref{eq:pr-Q} and introducing $\rho^{Q}_{X^n B^n} = \frac{1}{|T_Q^n|}\sum_{\bx\in T_Q^n} \ket{\bx}\bra{\bx}\otimes  \rho_{\bx}^{B^n}$ as the source of constant type $Q$ induced by $\rho_{XB}^{\otimes n}$. Note $\Pr(\by \in T_Q^n) > 0$ for each $Q\in \mathscr{P}_n(\cX^n)$ since $p(\bx)>0$ for each $\bx\in \cX^n$. From the code $\cC\source$  we will define codes $\cC\source^{(Q)} = (\cE\source^{(Q)}, \cD\source^{(Q)})$ for each $\rho_{X^nB^n}^{(Q)}$. We may define the encoder simply by restriction: $\cE^{(Q)} = \restr{\cE}{T_Q^n}$. This encoder has rate at most $R$.

The i.i.d.~source decoder is given by a family of POVMs, $\cD\source = \{ \{\Pi^{(z)}_\bx \}_{\bx\in \cX^n} \}_{z\in\cZ}$. We define the type~Q source decoder by restricting the POVMs:
\[
\cD\source^{(Q)} = \{ \{\Pi^{(z)}_\bx \}_{\bx\in T_Q^n} \}_{z\in\cZ}.
\]
Then
\begin{align}
\Pe(\cC\source^{(Q)}) = 1 - \frac{1}{|T_Q|^n}\sum_{\bx\in T_Q^n} \tr[ \Pi^{(\cE\source^{(Q)})}_\bx \rho^{B^n}_\bx ].
\end{align}
and $\sum_{Q\in \mathscr{P}_n(\cX^n)}\Pr(\by\in T_Q^n) \Pe(\cC\source^{(Q)})  = \Pe(\cC\source)$. By \eqref{eq:bound-Pstarchan-by-type-code} we have that for each $Q\in \mathscr{P}_n(\cX^n)$,
\begin{align}
\Pestarchan(n, H(Q)-R-\delta_n, Q)&\leq \left( 1 + \frac{1}{n} \right)\Pe(\cC\source^{(Q)})
\end{align}
so in particular,
\begin{align}
\Pestarchan(n, H(Q)-R-\delta_n, Q) \Pr(\by\in T_Q^n)  &\leq \left( 1 + \frac{1}{n} \right) \Pr(\by\in T_Q^n) \Pe(\cC\source^{(Q)}) \\
&\leq \left( 1 + \frac{1}{n} \right)\sum_{Q\in\mathscr{P}_n(\cX^n)}\Pr(\by\in T_Q^n) \Pe(\cC\source^{(Q)}) \\
&=  \left( 1 + \frac{1}{n} \right)\Pe(\cC\source).
\end{align}
Since the right-hand side no longer depends on $Q$, we may maximize over $Q$ to find
\begin{equation}
\max_{Q\in\mathscr{P}_n(\cX^n)} \Pestarchan(n, H(Q)-R-\delta_n, Q) \Pr(\by\in T_Q^n)   \leq \left( 1 + \frac{1}{n} \right)\Pe(\cC\source).
\end{equation}
Since this holds for any source code $\cC\source$ of rate $R$, we may minimize over such codes yielding
\begin{equation}
\max_{Q\in\mathscr{P}_n(\cX^n)} \Pestarchan(n, H(Q)-R-\delta_n, Q) \Pr(\by\in T_Q^n)   \leq \left( 1 + \frac{1}{n} \right) \Pestarsource(n,R).
\end{equation}
Thus,
\begin{align}
\min_{Q\in\mathscr{P}_n(\cX^n)} \left[- \frac{1}{n}\log(\Pr(\by\in T_Q^n)) + e\chan(n,H(Q)-R-\delta_n, Q) \right]\geq e\source(n,R) - \frac{1}{n}\log(1 + \frac{1}{n}).
\end{align}
Using \eqref{eq:prob_T},
\begin{align}
\frac{|\cX|}{n}\log(n+1) + \min_{Q\in\mathscr{P}_n(\cX^n)} \left[D(Q\|P)  + e\chan(n,H(Q)-R-\delta_n, Q) \right]\geq e\source(n,R) - \frac{1}{n}\log(1 + \frac{1}{n}).
\end{align}
which yields \eqref{eq:duality_op2}.

\section{Entropic duality} \label{sec:entropic_duality}

Theorems~\ref{thm:e-duality-constant-type} and \ref{thm:e-duality} show the duality between the type-dependent CQSW and the i.i.d.~CQSW with the classical-quantum channel coding with fixed composition on the operational level, i.e., in terms of the operational exponents  $e\chan(n,R,Q)$, $e\source(n,R,Q)$, $e\chan(n,R)$, and $e\source(n,R)$. One main focus of information theory is to bound operational quantities in terms of entropic quantities, which are entropic error exponents in this case. These are closed-form expressions which appear in bounds on the operational exponents. This naturally leads to the question of whether or not the relationships between $e\source$ and $e\chan$ investigated in Section~\ref{sec:operational_duality} persist at the level of the entropic quantities which bound them, i.e. if so-called ``entropic dualities'' hold. That is the main focus of this section.

In Section~\ref{sec:ent-dual-def}, we recall some necessary definitions before discussing entropic duality on the level of types in Section~\ref{sec:ent-dual-types}, in the i.i.d.~case in Section~\ref{sec:ent-dual-iid}, and in the strong converse case in Section~\ref{sec:ent-dual-sc}. In Section~\ref{sec:proof_entropic_duality}, we prove the main results entropic duality results.

\subsection{Definitions}\label{sec:ent-dual-def}

For any pair of density operators $\rho$ and $\sigma$, we define the {quantum relative entropy} is given by
\begin{equation}
D(\rho\|\sigma) :=  \Tr \left[ \rho \left( \log \rho - \log \sigma \right) \right]. \label{eq:relative}
\end{equation}
Given a bipartite density operator $\rho_{AB}$, its conditional entropy is defined by
\begin{equation} \label{eq:def_CE}
H(A|B)_\rho := - D (\rho_{AB}\|\one_A\otimes \rho_B). 
\end{equation}
In this work, families of divergences which generalize the quantum relative entropy play an important role: Petz's quantum R\'enyi divergence \cite{Pet86}, given by
\begin{equation}
D_\alpha(\rho\|\sigma) := \frac{1}{\alpha-1} \log K_\alpha(\rho\|\sigma) , \quad
K_\alpha(\rho\|\sigma) := \Tr \left[ \rho^\alpha \sigma^{1-\alpha} \right] \label{eq:Petz}
\end{equation}
and the sandwiched R\'enyi divergence \cite{MDS+13, WWY14}, given by
\begin{equation}
D^*_\alpha(\rho\|\sigma) := \frac{1}{\alpha-1} \log K^*_\alpha(\rho\|\sigma), \quad
K^*_\alpha(\rho\|\sigma) := \Tr \left[ \left( \rho^{\frac12} \sigma^{\frac{1-\alpha}{\alpha}} \rho^{\frac12} \right)^{\!\alpha}\,\right]. \label{eq:sandwich}
\end{equation}
Both quantities reduce to the quantum relative entropy in the limit $\alpha\to 1$. We use the notation $D_\alpha^t(\rho\|\sigma)$ to stand for either of the two entropies: $t=\{\}$ designates Petz's quantum R\'enyi divergence, while $t=\{*\}$ designates the sandwiched R\'enyi divergence. 

Given a c-q channel $\mathscr{W}:\mathcal{X}\to \mathcal{S}(B)$ and a prior probability $Q\in\mathscr{P}(\mathcal{X})$, we define
\begin{align}
I_0(Q,\mathscr{W}) &:= -\sup_{\tau_B \in \mathcal{S}(B)} \sum_{x\in\mathcal{X}} Q(x)  \log \Tr\left[ \Pi_{W_x} \tau_B\right],\\
I(Q,\mathscr{W}) &:= \sum_{x\in\mathcal{X}} Q(x) D\left(W_x \left\| Q\mathscr{W} \right. \right),
\end{align}
where $\Pi_{W_x}$ is the projection onto the support of $W_x := \mathscr{W}(x)$, and $Q\mathscr{W} := \sum_{{x} \in \mathcal{X}} Q({x})W_{{x}}$.
Similarly, given a set of states $\{ \rho_B^x \}_{x\in\mathcal{X}}$ and a prior probability $Q\in\mathscr{P}(\mathcal{X})$, we define
\begin{align}
\hat{H}_0(Q|B)_\rho &:= H(Q) + \sup_{\tau_B \in \mathcal{S}(B)} \sum_{x\in\mathcal{X}} Q(x)  \log \Tr\left[ \Pi_{\rho_B^x} \tau_B\right], \\
H(Q|B)_\rho &:= H(Q) - \sum_{x\in\mathcal{X}} Q(x) D\left( \rho_B^x \left\|\rho_B^Q \right. \right).
\end{align}
where $\rho_B^Q := \sum_{x\in\mathcal{X}} Q(x) \rho_B^x$. 

\subsection{Entropic duality on the level of types}\label{sec:ent-dual-types}
For $t=\{\}$ or $t=\{*\}$,  $s>-1$, we define the \emph{auxiliary function with type $Q$} associated to a finite set $\{\rho_B^x\}_{x\in \cX}\subset S(B)$ as
\begin{equation} \label{eq:E_0_Q0}
  E_0^t (s,Q)\equiv E_0^t (s,Q, \{\rho_B^x\}_{x\in \cX}) := s \Big[\inf_{ \tau_B \in \mathcal{S}(B) } \sum_{x\in\mathcal{X}} Q (x) D_{\frac{1}{1+s}}^t \left( \rho_B^{x} \| \tau_B \right) - H(Q)\Big].
\end{equation}
We also need another version of this quantity which does not have the minimization over $\tau_B$:
\begin{equation} \label{eq:E_0_Q0_downarrow}
  E_0^\downarrow (s,Q)\equiv E_0^\downarrow (s,Q, \{\rho_B^x\}_{x\in \cX}) := s \Big[ \sum_{x\in\mathcal{X}} Q (x) D_{1-s} \big( \rho_B^x \big\| \rho_B^Q \big) - H(Q)\Big].
\end{equation}
These auxiliary functions are used to define entropic exponents for the source coding of $\rho_{X^nB^n}^{(Q)}$ (defined in \eqref{eq:source-of-type-Q}), and the constant-composition channel coding of the map $\mathscr{W} : x\mapsto \rho_B^x$. 
We define the type-dependent entropic random coding\footnote{These are used for achievability bounds proven by constructing random codes, which is where these quantities get their name.} exponents for $R\geq 0$ by
 \begin{equation}
{E}\rs(R, Q) := \max_{0\leq s\leq 1} \left\{  {E}_0^\downarrow (s, Q) + s R \right\}, \qquad {E}\rc(R, P) := \max_{0\leq s\leq 1} \left\{  {E}_0^\downarrow (s, P) + s (H(P) - R) \right\},
\end{equation}
where the subscript ``s'' denotes source and the subscript ``c'' denotes channel. Next, the type-dependent sphere-packing exponents\footnote{In the classical case, Shannon, Gallager, and Berlekamp \cite{SGB67} used a sphere-packing technique to bound error exponents using these quantities, which is where these quantities get their name.} are defined for $R\geq 0$ as
\begin{equation}
 E\sps(R,Q)   := \sup_{s\geq 0}\left\{ E_0(s,Q) + sR\right\}, \qquad  E\spc(R,P)   := \sup_{s\geq 0}\left\{ E_0(s,P) + s(H(P)-R)\right\}.
\end{equation}

From the definitions given above and a joint continuity result in \cite[Theorem 6]{CGH18}, we characterize the properties of the entropic quantities in the following Proposition~\ref{prop:region}.
\begin{prop}[Properties of the  entropic error exponent functions] \label{prop:region}
	Assume $|\mathcal{X}|<\infty$.
	Given a c-q channel $\mathscr{W} : \cX \to \mathcal{S}(B)$ with and a set of states $\left\{ \rho_B^x \right\}_{x\in\mathcal{X}}$, the following hold.
	\begin{enumerate}
		\item The maps $E_\textnormal{r,c}^\downarrow (\cdot,\cdot)$ and  $E_\textnormal{sp,c}(\cdot,\cdot)$ are jointly continuous on $[0,\infty]\times \mathscr{P}(\mathcal{X})$ and $(I_0(P,\mathscr{W}),\infty]\times \mathscr{P}(\mathcal{X})$.
		Given every $P\in\mathscr{P}(\mathcal{X})$, both $E_\textnormal{r,c}^\downarrow (\cdot,P)$ $E_\textnormal{sp,c}(\cdot,P)$ are convex and non-increasing on $[0,\infty]$.
		In particular, for $P\in\mathscr{P}(\mathcal{X})$ with $I(P,\mathscr{W}) > I_0(P,\mathscr{W}) > 0$,
		\begin{align}
		E_\textnormal{r,c}^\downarrow (R,P) \in
		\begin{dcases}
		(0,\infty) & R< I(P,\mathscr{W}) \\
		\{0\}, & R\geq I(P,\mathscr{W})	
		\end{dcases}; \quad
		E_\textnormal{sp,c} (R,P) \in
		\begin{dcases}
		\{\infty\} & R< I_0(P,\mathscr{W}) \\
		(0,\infty) & R \in \left( I_0(P,\mathscr{W}), I(P,\mathscr{W}) \right)\\
		\{0\}, & R\geq I(P,\mathscr{W})	
		\end{dcases}.
		\end{align}
		
		\item 
		The maps $E_\textnormal{r,s}^\downarrow (\cdot,\cdot)$ and  $E_\textnormal{sp,s}(\cdot,\cdot)$ are jointly continuous on $[0,\infty]\times \mathscr{P}(\mathcal{X})$ and $[0,\hat{H}_0(Q|B)_\rho )\times \mathscr{P}(\mathcal{X})$.
		Given every $Q\in\mathscr{P}(\mathcal{X})$, both $E_\textnormal{r,s}^\downarrow (\cdot,Q)$ $E_\textnormal{sp,s}(\cdot,Q)$ are convex and non-decreasing on $[0,\infty]$.
		In particular, for $Q\in\mathscr{P}(\mathcal{X})$ with $\hat{H}_0(Q|B)_\rho > H(Q|B)_\rho > 0$,
		\begin{align}
		E_\textnormal{r,s}^\downarrow (R,Q) \in
		\begin{dcases}
		\{0\} & R\leq H(Q|B)_\rho \\
		(0,\infty], & R> H(Q|B)_\rho
		\end{dcases}; \quad
		E_\textnormal{sp,s} (R,Q) \in
		\begin{dcases}
		\{0\} & R\leq H(Q|B)_\rho \\
		(0,\infty) & R \in ( H(Q|B)_\rho, \hat{H}_0(Q|B)_\rho )\\
		\infty, & R > \hat{H}_0(Q|B)_\rho
		\end{dcases}.
		\end{align}
	\end{enumerate}
\end{prop}

These quantities bound the operational error exponents, as shown by the following result.
\begin{theo}[Constant composition channel coding bounds. Upper: \cite{DW14,CHT17}. Lower: Appendix~\ref{sec:chan-type-achiev}] \label{theo:entropic-chan-constant-type}
 Given a c-q channel $\mathscr{W} : \cX \to \mathcal{S}(B)$ and a type $Q\in \mathscr{P}_n(\cX)$, we have the following bounds on the exponent characterizing the optimal probability of error using codes of constant composition $Q$, $e\chan(n,R,Q)$: there exists positive constants $K_1,K_2$ and $N_0$ depending on $\mathscr{W}$, $|\mathcal{X}|$, and $R$ such that 
\begin{equation}
E_\textnormal{r,c}^{\downarrow}(R,Q) - \frac{K_1\log n}{n} \leq  e\chan(n,R,Q) \leq E_\textnormal{sp,c}(R,Q) + \frac{K_2\log n}{n}
\end{equation}
where 
the lower bound holds for all $R> 0$ and $n\in\mathbb{N}$, and the upper bound holds for $R \in (I_0(Q,\mathscr{W}), I(Q,\mathscr{W}) )$ and all $n\geq N_0$.
\end{theo}
\begin{remark}
The upper bound in Theorem~\ref{theo:entropic-chan-constant-type} was proven in the asymptotic regime by Dalai and Winter \cite{DW14}, i.e.~$\lim_{n\to \infty} e_\text{c}(n,R,Q) \leq E_\text{sp,c}(R,Q)$, and for finite blocklength by Ref.~\cite{CHT17}.
The lower bound is novel to our best knowledge, and is proven in Appendix~\ref{sec:chan-type-achiev}. We note that Hayashi \cite{Hay07} also proved an achievability lower bound for random codes with an i.i.d.~ensemble. 
Setting $Q\in\mathscr{P}_n(\mathcal{X})$ as the distribution of the i.i.d.~ensemble and the composition of the codes, we remark that our result, the lower bound of Theorem~\ref{theo:entropic-chan-constant-type}, is tighter than that of \cite{Hay07} by Jensen's inequality, i.e.~
\begin{align}
E_\text{r,c}^\downarrow(Q,R) &= \max_{0\leq s\leq 1} \left\{  {E}_0^\downarrow (s, P) + s (H(P) - R) \right\} \\
&= \max_{0\leq s\leq 1} \left\{ s \sum_{x\in\mathcal{X}} Q(x) D_{1-s}\left( \rho_B^x  \left\| \rho_B^Q \right. \right) - s R  \right\} \\
&= \max_{0\leq s\leq 1} \left\{ - \sum_{x\in\mathcal{X}} Q(x) \log \left[ K_{1-s} \left( \rho_B^x  \left\| \rho_B^Q \right. \right) \right] - s R  \right\} \\
&\geq \max_{0\leq s\leq 1} \left\{ -  \log  \left[ \sum_{x\in\mathcal{X}} Q(x)    K_{1-s} \left( \rho_B^x  \left\| \rho_B^Q  \right. \right) \right]- s R  \right\},
\end{align}
where the last line is the exponent obtained by Hayashi \cite{Hay07}.
We refer the reader to Ref.~\cite{Gal94} for the discussion of the achievability of i.i.d.~random codes and constant composition codes.
\end{remark}
The operational duality on the level of types given in Theorem~\ref{thm:e-duality-constant-type} therefore implies the following bounds on the source coding error exponents $e\source(n,R)$ and $e\source(n,R,Q)$.

 \begin{theo}[Error exponents for source coding of constant type] \label{theo:entropic_source_bounds-constant-type}
We have the following bounds on the exponent characterizing the optimal probability of error in the source coding of the source of constant type $\rho^{(Q)}_{X^n B^n}$,  $e\source(n,R,Q)$: there exists positive constants $K_1, K_2$ and $N_0$ depending on $\{\rho_B^x\}_{x\in\mathcal{X}}$, $|\mathcal{X}|$, $R$ such that
\begin{align}
e\source(n,R+\delta_n',Q) &\geq E_\textnormal{r,s}^{\downarrow}(R,Q) - \frac{K_1\log n}{n} - \frac{1}{n}, \label{eq:exp_s1}\\
e\source(n,R,Q) &\leq E_\textnormal{sp,s}(R+\delta_n',Q) + \frac{K_2\log n}{n} + \frac{1}{n}\log(1+\frac{1}{n}), \label{eq:exp_s2}
 \end{align}
where 
the lower bound holds for $R\in ( 0, H(Q) )$ and all $n\in\mathbb{N}$; the upper bound holds for  $R\in ( H(Q|B)_\rho, \hat{H}_0(Q|B)_\rho )$ and all $n\geq N_0$;
 $\delta_n$ and $\delta_n'$ are defined in \eqref{eq:def_delta_deltaprime}.
\end{theo}
The proof follows immediately from Theorem~\ref{theo:entropic-chan-constant-type}  and Theorem~\ref{thm:e-duality-constant-type}. 
Together, Theorems~\ref{theo:entropic-chan-constant-type} and \ref{theo:entropic_source_bounds-constant-type} give $E\rs$, $E\sps$ and $E\rc$, $E\spc$ operational motivations. In light of that, the relations
\begin{equation}
E\rs(R,Q) = E\rc(R,H(Q) - R), \qquad E\sps(R,Q) = E\spc(R,H(Q) - R), 
\end{equation}
which are immediate from the definitions, earn the designation of ``entropic duality on the level of types.''

\subsection{Entropic duality for i.i.d.~sources}\label{sec:ent-dual-iid}

In this case, we define the \emph{type-independent auxiliary function} associated to a c-q state $\rho_{XB}$ by
\begin{equation} \label{eq:E0SW1}	
E_0^t(s ) \equiv E_0^t(s, \rho_{XB}) := s \inf_{ \tau_B \in \mathcal{S}(B) }D_{\frac{1}{1+s}}^t \left( \rho_{XB} \| \one_A \otimes \tau_B \right),
\end{equation}
and the corresponding ``downarrow'' version by
\begin{equation}
E_0^\downarrow(s) \equiv E_0^\downarrow(s, \rho_{XB})  := s D_{1-s}(\rho_{XB}\|\one_X \otimes \rho_B).
\end{equation}
Then we define the type-independent random coding exponent,
\begin{equation}
 {E}\rs(R) := \max_{0\leq s\leq 1} \left\{  {E}_0^\downarrow (s ) + s R \right\}, 
\end{equation}
and sphere-packing exponent,
\begin{equation}
 E\sps (R)   := \sup_{s\geq 0}\left\{ E_0(s ) + sR\right\} \label{eq:Esp}.
\end{equation}

In~\cite{CHDH-2018}, the present authors  established the source coding bounds
\begin{align}
e\source(n,R) &\geq E\rs(R) - \frac{2}{n},  &  &\forall\, R> H(X|B)_\rho &&\\
e\source(n,R) &\leq E_\textnormal{sp,s} (R) + \tilde K \frac{\log n}{n} + K'\frac{1}{n}, & &\forall\, R \in (H(X|B)_\rho,H_0^\uparrow(X|B)_\rho) &&
\end{align}
where $\tilde K$ and $K'$ are constants which depend on $\{\rho_B^x\}_{x\in \cX}$, $|\cX|$, and $R$, and
\begin{equation}
 H_0^\uparrow(X|B)_\rho = \sup_{\tau_B\in S(B)} \log \tr[ \Pi_{\rho_{XB}}  \one_X\otimes\tau_B]
 \end{equation} where $\Pi_{\rho_{XB}}$ is the projection onto the support of $\rho_{XB}$.

On the other hand, Theorem~\ref{thm:e-duality} relates $e\source(n,R)$ to $e\chan(n,R,P)$. Thus, Theorem~\ref{theo:entropic-chan-constant-type}, which bounds the constant-type quantity $e\chan(n,R,Q)$, can be used along with Theorem~\ref{thm:e-duality} to establish an entropic bound on the i.i.d.~source coding error exponent $e\source(n,R)$. We find the following result.
 \begin{theo}
The exponent characterizing the optimal probability of error of an i.i.d.~source $\rho_{XB}^{\otimes n}$,  $e\source(n,R)$, satisfies: for some positive constants $K_1, K_2$ and $N_0$ depending on $\{\rho_B^x\}_{x\in\mathcal{X}}$, $|\mathcal{X}|$, $R$, 
\begin{align}
e\source(n,R+\delta_n' + \delta_n) &\geq \min_{Q \in \mathscr{P}_n(\cX)} [ D(Q\|P) + E\rs(R,Q)] - \frac{K_1\log n}{n} - \delta_n',  \label{eq:exp_s3}\\
e\source(n,R) &\leq \min_{Q \in \mathscr{P}_n(\cX)} [ D(Q\|P) + E_{\textnormal{sp,s}}(R+ \delta_n',Q)] + \frac{K_2\log n}{n} + \delta_n',\label{eq:exp_s4}
\end{align}
where $P$ is the source distribution, namely $\{p(x)\}_{x\in \cX}$ for $\rho_{XB}^{\otimes n}$ as in \eqref{eq:iid-state}, and
 $\delta_n$ and $\delta_n'$ are defined in \eqref{eq:def_delta_deltaprime}.
Note that the lower bound holds for all $R>0$ and $n\in \mathbb{N}$, and the upper bound holds for $R \in (H(X|B)_\rho,H_0^\uparrow(X|B)_\rho)$ and all $n\geq N_0$.
 \end{theo}
This result motivates the definition of the entropic quantities
\begin{equation}
\hat E\rs(R) :=\min_{Q \in \mathscr{P}(\cX)} [ D(Q\|P) + E\rs(R,Q)] = \min_{Q \in \mathscr{P}(\cX)} [ D(Q\|P) + E\rc(H(Q) - R,Q)]
\end{equation}
and
\begin{equation} \label{eq:def_hat_Esps}
\hat E\sps(R) := \min_{Q \in \mathscr{P}(\cX)} [ D(Q\|P) + E\sps(R,Q)] = \min_{Q \in \mathscr{P}(\cX)} [ D(Q\|P) + E\spc(H(Q) - R,Q)]. 
\end{equation}
Note that these quantities are defined with a minimization over probability distributions $Q$ instead of over types.  Since
\begin{equation}
\hat E\rs(R) \leq \min_{Q \in \mathscr{P}_n(\cX)} [ D(Q\|P) + E\rs(R,Q)]
\end{equation}
the quantity $\hat E\rs(R)$ can be used for a lower bound on $e\source(n,R)$ via~\eqref{eq:exp_s3}. For the sphere-packing exponent, however, $\hat E\sps(R)$ does not immediately bound $e\source(n,R)$ at finite $n$ via \eqref{eq:exp_s4}; while we have $\limsup_{n\to\infty} e\source(n,R) \leq \hat E\sps(R)$ using the continuity results\footnote{Note $Q_n(x) = \frac{1}{n}\floor{Q(x) n} \in \mathscr{P}_n(\cX)$ gives an approximation of any distribution $Q$ in which the error $\|Q  - Q_n\|_\infty \leq \frac{1}{n}$ can be bounded independently of $Q$.} of Proposition~\ref{prop:region}, for a finite~$n$ result one requires an explicit continuity bound for $Q\mapsto E\sps(R,Q)$, which is an open problem.

This naturally leads to the question of comparison between $\hat E\rs(R)$ and $E\rs(R)$, and between $\hat E\sps(R)$ and $E\sps(R)$, which is partially resolved by the following result.
\begin{theo} \label{theo:Esp-minimax-duality}
We have for $R\geq 0$,
\begin{equation}
\hat E\sps(R) = E\sps(R),
\end{equation}
and moreover, for $s\geq 0$,
\begin{equation}
E_0 (s) = 
\min_{Q \in \mathscr{P}(\mathcal{X}) } \left\{D(Q\|P)  +  E_0(s,Q) \right\}
\end{equation}
where $P = \{p(x)\}_{x\in \cX}$ is the distribution of the i.i.d.~source.
\end{theo}
We defer the proof to Section~\ref{sec:proof_entropic_duality}.  Theorem~\ref{theo:Esp-minimax-duality} shows that for $R\geq 0$,
\begin{equation} \label{eq:sp-duality}
 E\sps(R) = \min_{Q \in \mathscr{P}(\cX)} [ D(Q\|P) + E\spc(H(Q) - R,Q)]
\end{equation}
by \eqref{eq:def_hat_Esps}. By analogy to Theorem~\ref{thm:e-duality}, Equation~\eqref{eq:sp-duality} can be regarded as a statement of entropic duality.

\subsection{Strong converse entropic duality} \label{sec:ent-dual-sc}
We define the type-dependent entropic strong converse exponents by
\begin{equation}\label{eq:sc-exponents-type}
E\scs (R, Q) := \sup_{-1 < s< 0}\left\{ {E}^*_0 (s, Q) + sR\right\}, \qquad E\scc (R, P) := \sup_{-1 < s< 0}\left\{ {E}^*_0 (s, P) + s(H(P) - R)\right\}.
\end{equation}
We also define the type-independent strong-converse entropic exponent for source coding by
\begin{equation}\label{eq:sc-exponent-iid}
E\scs (R) := \sup_{-1 < s< 0}\left\{ {E}^*_0 (s ) + sR\right\}.
\end{equation}
In the absence of an operational duality of the strong converse exponents, we cannot follow the same path as in the previous sections. Nonetheless, we can bound $sc\source(n,R,Q)$ using $E\scs(R,Q)$ and $sc\chan(n,R,Q)$ using $E\scc(R,Q)$ independently, as well as bound $sc\source(R)$ using $E\scs(R)$, giving each of the entropic quantities defined in \eqref{eq:sc-exponents-type} and \eqref{eq:sc-exponent-iid} operational motivation. This is done in Appendix~\ref{sec:sc_proofs}, yielding the following result.
\begin{prop}[Entropic bounds on the strong converse exponents]
For all $R>0$ and $n\in\mathbb{N}$:
\begin{align} \label{eq:sc_LB}
sc\source(n,R,Q) &\geq E_\textnormal{sc,s}^* (R, Q) - |\mathcal{X}| \frac{\log(n+1)}{n},\\
\label{eq:sc_LB1}
sc\chan(n,R,Q) &\geq E_\textnormal{sc,c}^*(R,Q), \\
sc\source(n,R) &\geq E_\textnormal{sc,s}^*(R) - 2|\mathcal{X}| \frac{\log(n+1)}{n}. \label{eq:E_0_Q3_sc}
\end{align}
In particular, $E_\textnormal{sc,s}^* (R, Q)>0$ for $R<H(Q|B)_\rho$, $E_\textnormal{sc,c}^*(R,Q)>0$ for $R>I(Q,\mathscr{W})$, and $E_\textnormal{sc,s}^*(R)>0$ for $R< H(X|B)_\rho$.
\end{prop}
\begin{remark}
In \cite{CHDH-2018}, the present authors established a tighter bound on $sc\source(n,R)$: for all $R < H(X|B)_\rho$ and $n\in \mathbb{N}$,
\begin{equation}
sc\source(n,R) \geq E_\textnormal{sc,s}^*(R)
\end{equation}
along with an asymptotically matching upper bound, using different techniques. We include \eqref{eq:E_0_Q3_sc} to show how the type-dependent bound \eqref{eq:sc_LB1} can quickly yield type-independent results.
\end{remark}
In light of these bounds, the mathematical identity
\begin{equation} \label{eq:sc-duality-type}
E\scs(R,Q) = E\scc(H(Q)-R,Q)
\end{equation}
is an ``entropic duality for the strong converse exponents'' on the level of types. In fact, $E\scs(R)$ and $E\scs(R,Q)$ are related in the same manner as $E\sps(R)$ and $E\sps(R,Q)$.
\begin{theo} \label{theo:Esc-minimax-duality}
For $R\geq 0$, we have the identity
\begin{equation}
E\scs(R) = \min_{Q \in \mathscr{P}(\mathcal{X}) } \left\{D(Q\|P) + E\scs(R, Q) \right\}
\end{equation}
and in fact, for $s\in(-1,0)$
\begin{equation}
E_0^* (s) = \min_{Q \in \mathscr{P}(\mathcal{X}) } \left\{D(Q\|P) + E_0^*(s,Q) \right\}
\end{equation}
where $P = \{p(x)\}_{x\in \cX}$ is the distribution of the i.i.d.~source.
\end{theo}
We defer the proof to the following section. Using \eqref{eq:sc-duality-type}, Theorem~\ref{theo:Esc-minimax-duality} yields
\begin{equation}
E\scs(R) = \min_{Q \in \mathscr{P}(\mathcal{X}) } \left\{ D(Q\|P) + E\scc(H(Q)-R,Q)\right\},
\end{equation}
where $E\scs$ is defined with respect to the c-q channel $\cX\ni x\mapsto \rho_B^x$, which can be regarded as a strong converse entropic duality between i.i.d.~source coding and constant composition c-q channel coding, similar to \eqref{eq:sp-duality}.

\subsection{Proof of Theorems \ref{theo:Esp-minimax-duality} and \ref{theo:Esc-minimax-duality}}\label{sec:proof_entropic_duality}

The proofs of Theorems \ref{theo:Esp-minimax-duality} and \ref{theo:Esc-minimax-duality} rely heavily on the following classical minimax theorem.
\begin{lemm}[Ky Fan \cite{fan_minimax_1953}]
\label{lemm:minimax} If $X$ is a compact convex set in a topological vector space $V$, $Y$ is a convex subset of a vector space $W$, and $f: X\times Y\to \mathbb{R}$ has the properties that 
\begin{enumerate}
	\item $y \mapsto f(x,y)$ is concave and upper-semicontinuous on $Y$ for each $x\in X$,
	\item $x\mapsto f(x,y)$ is convex and lower-semicontinuous on $X$ for each $y\in Y$
\end{enumerate}
then 
\begin{equation}
\min_{x\in X} \sup_{y\in Y} f(x,y) = \sup_{y\in Y} \min_{x\in X} f(x,y)
\end{equation}
where the supremum may be replaced by a maximum in the case that $Y$ is compact.
\end{lemm}
We also use the following lemma to establish the concavity of the auxiliary functions.
\begin{lemm}[Concavity of the auxiliary functions \cite{CGH18,MO17}] \label{lemm:aux_concave}
	Let $\left\{\rho_B^x\right\}_{x\in\mathcal{X}}$ be a set of density operators in $\mathcal{S}(B)$. For any $Q \in \mathscr{P}(\mathcal{X})$, the maps
	\begin{align}
	(-1,0) \ni s \mapsto E_0^*(s,Q) \label{eq:aux-concave-sc}\\
	[0,\infty) \ni s \mapsto E_0(s,Q) \label{eq:aux-concave-sp}
	\end{align}
	are concave.
\end{lemm}
\begin{remark}
Equation~\eqref{eq:aux-concave-sc} is due to \cite{CGH18}, while Equation~\eqref{eq:aux-concave-sp} is Cor. B.2 of \cite{MO17}.
\end{remark}
Lastly, we employ the following variational formula characterizing the classical Renyi relative entropy as a tradeoff between two relative entropy terms.
\begin{lemm}[{\cite[Theorem 30]{EH14}}] \label{lemm:class-var}
	Let $P$ and $Q$ be probability distributions on $\cX$.
	For all $s>-1$, it follows that
	\begin{align}
	\min_{R \in \mathscr{P}(\cX)} D(R\|P) + sD(R\|Q)
	= s D_{\frac{1}{1+s}}(P\|Q).
	\end{align}
\end{lemm}
\begin{remark}
See \cite[Theorem III.5]{MO17} for a quantum generalization of this result.
\end{remark}
For the convenience of the reader, we summarize Theorems~\ref{theo:Esp-minimax-duality} and \ref{theo:Esc-minimax-duality} as the following proposition.

\begin{prop}[Entropic duality] \label{prop:E_0_Q}
Let $\rho_{XB} = \sum_{x\in\mathcal{X}} p(x) |x\rangle\langle x|\otimes \rho_B^x$ be a c-q state, and $P =\{p(x)\}_{x\in \cX}$. For $R\geq 0$, we have
	\begin{align}
	E_\textnormal{sp,s}(R) &= \min_{Q \in \mathscr{P}(\mathcal{X}) }  \left\{  D(Q\|P) + E_\textnormal{sp,s}(R,Q)  \right\}. \label{eq:duality_Esp}\\
	E_\textnormal{sc,s}^*(R) &= \min_{Q \in \mathscr{P}(\mathcal{X}) } \left\{ D(Q\|P) + E_\textnormal{sc,s}^*(R, Q)  \right\} \label{eq:duality_Esc}.
	\end{align}
	In fact, the auxiliary functions exhibit a duality as well. For $s\geq 0$,
	\begin{equation} \label{eq:E_0_Q}
	E_0 (s) = 
	\min_{Q\in \mathscr{P}(\mathcal{X}) } \left\{ D(Q\|P) + E_0(s,Q)  \right\}, 
	\end{equation}
	and for $s\in(-1,0)$,
	\begin{equation} \label{eq:E_0_Q_star}
	E_0^* (s) = \min_{Q \in \mathscr{P}(\mathcal{X}) } \left\{ D(Q\|P)  + E_0^*(s,Q) \right\},
	\end{equation}
	where for $t=\{\}, \{*\}$, the quantity $E_0^t(s,Q)$ is introduced in Eq.~\eqref{eq:E_0_Q0} and
	$E_0^t(s)$ is defined by \eqref{eq:E0SW1}.
\end{prop}

\begin{remark}
	It is worth emphasizing that the entropic duality established in Proposition~\ref{prop:E_0_Q} is even new in the classical case.
	As $R \in [H(X|Y), H_{1/2}(X|Y)]$ (here the side information is denoted by the classical system $Y$), the error exponents of the i.i.d.~sources and the constant composition channel coding are determined respectively by $E_\text{sp,s}(R)$ and $E_\text{sp,c}(R,Q)$ \cite{Gal76,Kos77, CK80, CK81, Csi82, , Csi98, CK11}. Via the classical operational duality proved by Ahlswede and Dueck \cite{AD82}, Eq.~\eqref{eq:duality_Esp} can be shown for $R \in [H(X|Y), H_{1/2}(X|Y)]$.
	On the other hand, since the operational duality is unknown for $R \in [0,H(X|Y)]$, Eq.~\eqref{eq:duality_Esc} is new in the classical case.
	
\end{remark}

The proof of Proposition~\ref{prop:E_0_Q} employs the following properties of entropic quantities, proven in \cite{MO17}:
\begin{enumerate}
	\item \label{it:1} For $\rho, \sigma \in\mathcal{S}(B)$, the map $\alpha\mapsto D_\alpha^t(\rho\|\sigma)$ is continuous on $[0,\infty]$, for $t=\{\}$ and for $t=\{*\}$. (Cor. III.14)
	\item \label{it:2} For $\rho \in \mathcal{S}(B)$ and $s > -1$, the map $\sigma\mapsto D^*_{\frac{1}{1+s}}(\rho \| \sigma)$ is convex on $\mathcal{S}(B)$. (Prop. III.17)
	\item \label{it:3} For $s\geq 0$,  $\tau_B\to K_{\frac{1}{1+s}}(\rho_B^x\|\tau_B)$ is concave on $\mathcal{S}(B)$ (Prop. III.17).
	\item \label{it:4} For $s\in (-1,0)$ and $\rho\in S(B)$, $\sigma\mapsto D^t_{\frac{1}{1+s}}(\rho\|\sigma)$ is lower semicontinuous on $S(B)$. (Cor. III.26)
\end{enumerate}

\begin{proof}
Define	
\begin{align} \label{eq:def_G}
G^t(s,Q,\tau) := (1+s) D( Q \| \{ P(x)^{\frac{1}{1+s}} K_{\frac{1}{1+s}}^t(\rho_B^x \|\tau_B) \}).
\end{align}
Since the relative entropy is lower semicontinuous and jointly convex, we see that for $s > -1$, the map $Q\mapsto G^t(s,Q,\tau)$ is lower semicontinuous and convex. We may rewrite this quantity in terms of
\begin{equation}
P^{(s,\tau)}(x) := P(x) K_{\frac{1}{1+s}}^t(\rho_B^x\|\tau_B)^{1+s}
\end{equation}
yielding
\begin{equation} \label{eq:G-as-sum-of-entropies}
G^t(s,Q,\tau) = D(Q\| P^{(s,\tau)} ) + s D(Q\|\one).
\end{equation}
Then
\begin{align}
G^t(s, Q,\tau) &=- \sum_{x\in\mathcal{X}} Q (x) \log \big(K_{\frac{1}{1+s}}^t(\rho_B^x\|\tau_B)^{1+s}\big) + D(Q\|P) + s D\left( Q \| \mathds{1} \right) \\
&= s  \sum_{x\in\mathcal{X}} Q (x) D_{\frac{1}{1+s}}^t \left( \rho_B^{x} \| \tau_B \right) - s H(Q) + D(Q\| P). \label{eq:G_as_relent}
\end{align}
From this expression, we see $s\mapsto G^t(s,Q,\tau)$ is continuous for all $s > -1$, using Property~\ref{it:1}, and that for $s\in(-1,0)$, the map $\tau \mapsto G^*(s,Q,\tau)$ is concave by Property~\ref{it:2}, and upper semicontiuous by Property~\ref{it:4}.
From this expression, we also find
\begin{equation} \label{eq:E_0Q_as_G}
E_0^t(s, Q) + D(Q\|P) = \begin{cases}
\min_{\tau \in S(B)} G^t(s,Q,\tau) & s\geq 0 \\
\max_{\tau \in S(B)} G^t(s,Q,\tau) & s\in(-1,0)
\end{cases}
\end{equation}
so the right-hand side of \eqref{eq:E_0_Q} is given by
\begin{equation} \label{eq:RHS_G_Q}
\begin{cases}
\min_{Q\in \mathscr{P}(\cX)}\min_{\tau \in S(B)} G^t(s,Q,\tau) & s\geq 0 \\
\min_{Q\in \mathscr{P}(\cX)} \max_{\tau \in S(B)} G^t(s,Q,\tau) & s\in(-1,0)
\end{cases}
\end{equation}
On the other hand, we may use the form \eqref{eq:G-as-sum-of-entropies} to optimize $G^t$ over $Q$ using Lemma~\ref{lemm:class-var}. We find
\begin{align}
\min_{Q\in \mathscr{P}(\cX)} G^t(s,Q,\tau) &=\min_{Q \in \mathscr{P}(\mathcal{X}) } \left\{   D\left( Q \| P^{(s,\tau)}\right) + 	s D\left( Q \| \mathds{1} \right) 
\right\}  \\
&= s D_{\frac{1}{1+s}} \left( P^{(s,\tau)} \| \mathds{1} \right)\\
&= -(1+s) \log \sum_{x\in\mathcal{X}} (P^{(s,\tau)})^{\frac{1}{1+s}}(x) \\
&= -(1+s) \log \sum_{x\in\mathcal{X}} {P}^{\frac{1}{1+s}}(x) K^t_{\frac{1}{1+s}}(\rho_B^x\|\tau_B)  \\
&= s D_{\frac{1}{1+s} }^t \left( \rho_{XB} \| \mathds{1} \otimes  \tau_B \right).
\end{align}
Thus,
\begin{equation} \label{eq:E_0_as_G}
E_0^t(s) = \begin{cases}
\min_{\tau_B\in\mathcal{S}(B)}  \min_Q G^t(s,Q,\tau) & s \geq 0 \\
\max_{\tau_B\in\mathcal{S}(B)}  \min_Q G^t(s,Q,\tau) & s \in(0,-1).
\end{cases}
\end{equation}

Therefore, since we can exchange minima, \eqref{eq:E_0_as_G} and \eqref{eq:RHS_G_Q} with $t=\{\}$ and $s\geq 0$ yield \eqref{eq:E_0_Q}. On the other hand, for $s\in (0,-1)$ and $t=\{*\}$ we have established  $\tau \mapsto G^*(s,Q,\tau)$ is concave and upper semicontinuous, and $Q\mapsto G^*(s,Q,\tau)$
is convex and lower semicontinuous. Thus, the minimax result of Lemma~\ref{lemm:minimax} allows the interchange the minimum and maximum in \eqref{eq:E_0_as_G} $s\in (0,-1)$ and $t=\{*\}$. Comparison with \eqref{eq:RHS_G_Q} then yields \eqref{eq:E_0_Q_star}.

Next, for $R\geq 0$ define
\begin{equation} \label{eq:H1}
H^t(s,Q) = \begin{cases}
\min_{\tau_B \in \mathcal{S}(B)} G^t(s,Q,\tau) + s R  & s\geq 0\\
\max_{\tau_B \in \mathcal{S}(B)} G^t(s,Q,\tau) + s R  & s\in(-1,0).
\end{cases}
\end{equation}
then we have
\begin{equation} \label{eq:H-as-aux}
E_0^t(s,Q) + D(Q\|P)  +sR =  H^t(s,Q)
\end{equation}
by \eqref{eq:E_0Q_as_G}.
In particular, \eqref{eq:duality_Esp} is equivalent to
\begin{equation} \label{eq:minimax-Esp}
\sup_{s\geq 0} \min_{Q\in \mathscr{P}(\cX)} H(s,Q) = \min_{Q\in \mathscr{P}(\cX)} \sup_{s\geq 0} H(s,Q)
\end{equation}
while \eqref{eq:duality_Esc} is equivalent to
\begin{equation} \label{eq:minimax-Esc}
\sup_{s\in (-1,0)} \min_{Q\in \mathscr{P}(\cX)} H^*(s,Q) = \min_{Q\in \mathscr{P}(\cX)} \sup_{s\in (-1,0)} H^*(s,Q).
\end{equation}
It remains to check the assumptions of Lemma~\ref{lemm:minimax} in each case. We have that on the domain $s\in(-1,0)$ and on the domain $s \geq 0$, the map $s\mapsto H^t(s,Q)$ is continuous, as shown by the following argument. Recall $s\mapsto G(s,Q,\tau)$ is continuous. Then, since the set of density matrices is compact, $s \mapsto \min_{\tau_B\in\mathcal{S}(B)}  G(s,Q,\tau)$ and $s\mapsto \max_{\tau_B\in\mathcal{S}(B)}  G(s,Q,\tau)$ are continuous, by Cor. 3.1.22 of \cite{aubin_applied_1984}. Similarly, $Q\mapsto H^t(s,Q)$ is lower semicontinuous on the domain $s\in(-1,0)$ and on $s\geq 0$, using that $Q\mapsto G(s,Q,\tau)$ is lower semicontinuous, and Cor. 3.1.22 of \cite{aubin_applied_1984} again. Additionally, Lemma~\ref{lemm:aux_concave} gives that $s\mapsto H^*(s,Q)$ is concave for $s\in (-1,0)$ and $s\mapsto H(s,Q)$ is concave for $s\geq 0$, using \eqref{eq:H-as-aux}.

It remains to establish that $Q \mapsto H(s,Q)$ convex on $s\in [0,\infty)$, and $Q\mapsto H^*(s,Q)$ is convex on $s\in (-1,0)$.
For the latter, we note that $G^*(s,Q,\tau)$ is again convex in $Q\in \mathscr{P}(\mathcal{X})$ by Eq.~\eqref{eq:def_G} and the convexity of the relative entropy in its first argument. Then, from Eq.~\eqref{eq:H1}, we see that $H^*(s,Q) $ is convex in $Q\in \mathscr{P}(\mathcal{X})$ because pointwise supremum of convex functions is convex. This completes the proof of \eqref{eq:minimax-Esc}.

Lastly, we show $Q\mapsto H(s,Q)$ is convex by establishing the joint convexity of $(Q,\tau) \mapsto G(s,Q,\tau)$ for $s \geq 0$.
Let $Q_0, Q_1 \in \mathscr{P}(\mathcal{X})$, $\tau_0, \tau_1 \in \mathcal{S}(B)$, $\theta \in [0,1]$, $Q = (1-\theta) Q_0 + \theta Q_1$, and $\tau = (1-\theta) \tau_0 + \theta \tau_1$.
	 Then it follows that
	\begin{align}
	G(s, Q, \tau) &= (1+s) \sum_x Q(x) \log \frac{ Q(x) }{  P(x)^{\frac{1}{1+s}} K_{\frac{1}{1+s}}(\rho_B^x\|\tau)  } \label{eq:G1}\\
	&\leq (1+s) \sum_x Q(x) \log \frac{ Q(x) }{ (1-\theta)P(x)^{\frac{1}{1+s}} K_{\frac{1}{1+s}}(\rho_B^x\|\tau_0)  + \theta P(x)^{\frac{1}{1+s}} K_{\frac{1}{1+s}}(\rho_B^x\|\tau_1) } \label{eq:E_0_Q2} \\
	&\leq (1-\theta) (1+s) \sum_x Q_0(x) \log \frac{ Q_0(x) }{  P(x)^{\frac{1}{1+s}} K_{\frac{1}{1+s}}(\rho_B^x\|\tau_0)  } \nonumber\\
	&\qquad\qquad+ \theta (1+s) \sum_x Q_1(x) \log \frac{ Q_1(x) }{  P(x)^{\frac{1}{1+s}} K_{\frac{1}{1+s}}(\rho_B^x\|\tau_1)  } \label{eq:E_0_Q3} \\
	&= (1-\theta) G(s,Q_0, \tau_0) + \theta G(s,Q_1, \tau_1),
	\end{align}
	where inequality~\eqref{eq:E_0_Q2} follows from the concavity of $K_{\frac{1}{1+s}}(\rho_B^x\|\,\cdot\,)$, namely Property~\ref{it:3} above, and the monotonicity of the logarithm;
	and inequality~\eqref{eq:E_0_Q3} is due to the joint convexity of the relative entropy.
	Lastly, since $(Q, \tau_B)\mapsto G(s,Q, \tau_B)$ is jointly convex, it holds that $Q \mapsto \min_{\tau_B\in\mathcal{S}(B)} G(s,Q, \tau_B)$ is convex, which completes the proof of \eqref{eq:minimax-Esp}.
\end{proof}

\section{Discussions} \label{sec:conclusions}

We consider the task of classical source coding with quantum side information (QSI), and the task of c-q channel coding. The QSI can be considered to result from the action of a c-q channel on a copy of the output of the source. This naturally associates a c-q channel to the source coding task. We generalize the work of Ahlswede and Dueck \cite{AD82} to show that these two tasks exhibit a duality: good codes for one task can be used to make good codes for the other. This duality exists at the level of the \emph{error exponents}: the exponential rates of decay of the probability of error in $n$, which denotes the number of uses of the channel or source. 

In information theory, one typically finds expressions (or bounds) on operational quantities (e.g.\ optimal rates or error exponents) of an information-theoretic task in terms of entropic quantities. Error exponents for both classical source coding with QSI and c-q channel coding admit entropic upper bounds in terms of the so-called sphere-packing exponents, and entropic lower bounds in terms of the so-called random coding exponents. We show that the sphere-packing exponents for these two tasks satisfy an exact duality relation. Such a duality does not seem to be satisfied by the entropic random coding exponents, however, according to numerical results and preliminary analytic analysis. The apparent failure of duality in this case could indicate that the random coding exponents considered here are not optimal. In fact, in the case of source coding with QSI, it is believed that a tighter lower bound (achievability bound) on the error exponent may be possible; in~\cite{CHDH-2018}, the present authors \emph{conjectured} that a quantity denoted in that paper as $E_\text{r}(R)$ also yields an achievability bound for this task. Moreover, this quantity and its c-q channel analog satisfy an exact duality relation\footnote{The proof follows exactly as the proof in the sphere-packing case}.

A natural question to ask is whether the duality between classical source coding with QSI and c-q channel coding also manifests itself when considering the exponents associated to the probability of success (the strong converse exponents) instead of the probability of error. The strong converse exponents for both c-q channel coding and classical coding with QSI have exact entropic expressions in the asymptotic limit \cite{MO17,CHDH-2018}. We showed that these expressions, termed \emph{entropic strong converse exponents}, satisfy an exact duality relation. This shows that a mathematical duality holds for the asymptotic operational strong converse exponents, and may be suggestive of an underlying \emph{operational} strong converse duality, i.e.~one based on the construction of codes with finite $n$. The existence of such an operational duality is an interesting open question which likely requires new techniques.

\appendix

\section{Remaining proofs} \label{sec:app_proofs}

\subsection{Proof of Lemma~\ref{lem:covering}} \label{sec:proof-covering}
	The first statement, that there exist a set of $L_Q$ permutations which induces a cover of $T_Q^n$, is the type covering lemma of Ahlswede,  \cite[Section 5.6.1]{Ahl80}, which follows directly from the Covering Lemma of \cite[Section 2.3]{Ahl79}, which deals with a more general setting using the language of hypergraphs. We restate that proof, in the case of permutations and type classes, with a slight strengthening to yield both statements\footnote{We note that the same reasoning we use to strengthen the result in our setting holds in the general setting as well.}.

	Let $\bx\in T_Q^n$ be fixed and let $\pi$ be a permutation of $\{1,\dotsc,n\}$ taken uniformly at random. Then for a fixed $\bu \in \mathcal{U}$, the probability that $\pi(\bu) = \bx$ is $\frac{1}{|T_Q^n|}$ (since it has uniform probability of mapping $\bu$ to any sequence in $T_Q^n$). Then, since $\pi(\bu)= \bx$ implies $\bx = \pi\inv(\bu)$, we cannot have both $\pi(\bu) = \bx$ and $\pi(\bu') = \bx$ for $\bu \neq \bu' \in \mathcal{U}$. This mutual exclusivity implies equality in the union bound, namely
	\[
	\Pr(\bx \in \pi \mathcal{U}) = \sum_{\bu \in \mathcal{U}} \Pr(\bx = \pi(\bu)) = |\mathcal{U}|\cdot |T_Q^n|\inv,
	\]
	so $\Pr(\bx \not \in \pi\mathcal{U}) = 1 -|\mathcal{U}|\cdot |T_Q^n|\inv$.

	Next, choose $k$ permutations $\pi_1,\dotsc,\pi_k$ of $\{1,\dotsc,n\}$ independently and uniformly at random. Then, using independence,
	\[
	\Pr\left[ \bx \not \in \bigcup_{i=1}^k \pi_i \mathcal{U}\right] = \prod_{i=1}^k \Pr(\bx \not \in \pi_i \mathcal{U}) =  (1 - |\mathcal{U} |T_Q^n|\inv)^k.
	\]
	Thus, summing over $\bx\in T_Q^n$, the probability that there exists $\bx\in T_Q^n$ which is not covered is
	\[
	\Pr\left[ \exists \bx \in T_Q^n \text{ s.t. } \bx \not \in \bigcup_{i=1}^k \pi_i \mathcal{U}\right] \leq |T_Q^n| \cdot (1 - |\mathcal{U} |T_Q^n|\inv)^k
	\]
	using a union bound.

	Let $\delta \geq 0$. Then we have the equivalences
	\begin{gather}	
	|T_Q^n| \cdot (1 - |\mathcal{U} |T_Q^n|\inv)^k < 1 - \delta\\ 
	k \log [1 - |\mathcal{U}|\cdot |T_Q^n|\inv] < \log[|T_Q^n|\inv (1-\delta)]\\
	k > \frac{\log[|T_Q^n|\inv (1-\delta)]}{\log [1 - |\mathcal{U}|\cdot |T_Q^n|\inv]}. \label{ieq:final-k}
	\end{gather}
	Then, the inequality $\log(1+x) < x$, valid for all $x\in (-1,0)\cup(0,\infty)$, applied to $x = - |\mathcal{U}|\cdot |T_Q^n|\inv \in(-1,0)$ yields
	\begin{equation}
	- \log [1 - |\mathcal{U}|\cdot |T_Q^n|\inv] > |\mathcal{U}|\cdot |T_Q^n|\inv.
	\end{equation}
	Therefore, the inequality
	\begin{equation} 
		k \geq \frac{- |T_Q^n|\log[|T_Q^n|\inv (1-\delta)]}{|\mathcal{U}|}= \frac{|T_Q^n|(\log[|T_Q^n|]  - \log (1-\delta))}{|\mathcal{U}|}
		\label{eq:k-bound-delta}
	\end{equation}
	implies \eqref{ieq:final-k}.
	At this stage, we see if $\delta=0$, then for $k = L_Q := \ceil{|T_Q^n|\log[|T_Q^n|] |\mathcal{U}|\inv}$ we recover 
	\begin{equation}
	\Pr\left[ \exists\, \bx \in T_Q^n \text{ s.t. } \bx \not \in \bigcup_{i=1}^k \pi_i \mathcal{U}\right] \leq |T_Q^n| \cdot (1 - |\mathcal{U} |T_Q^n|\inv)^k < 1
	\end{equation}
	implying there must exist a set of $L_Q$ permutations which induce a cover of $T_Q^n$. 

	To prove the second statement, we consider $k = 2 L_Q$, and solve for $\delta$ in \eqref{eq:k-bound-delta}. We wish to find $\delta$ such that
	\[
	2 L_Q - |T_Q^n|\log[|T_Q^n|] |\mathcal{U}|\inv \geq - |T_Q^n|\cdot |\mathcal{U}|\inv \log(1-\delta).
	\]
	This is implied by
	\[
	|T_Q^n|\log[|T_Q^n|] |\mathcal{U}|\inv \geq- |T_Q^n|\cdot |\mathcal{U}|\inv \log(1-\delta)
	\]
	using the bound $L_Q \geq |T_Q^n|\log[|T_Q^n|] |\mathcal{U}|\inv$ for each $L_Q$. Therefore, if 
	\[
	|T_Q^n| = \frac{1}{1-\delta} \iff 	1 - \delta = \frac{1}{|T_Q^n|},
	\]
	then  \eqref{eq:k-bound-delta} holds for $k=2 L_Q$.	Thus,
	\[
		\Pr\left[ \exists \bx \in T_Q^n \text{ s.t. } \bx \not \in \bigcup_{i=1}^{2L_Q} \pi_i \mathcal{U}\right] < \frac{1}{|T_Q^n|}
	\]
	and so the probability of $\pi_1,\dotsc,\pi_{2L_Q}$ inducing a cover of $T_Q^n$ is at least $1- \frac{1}{|T_Q^n|}$.

	Next, we wish to find the expected number of trials (of independent draws of $2L_Q$ permutations) until we find a cover. This is governed by a geometric distribution with success probability $p = 1- \frac{1}{|T_Q^n|}$, and the expected number of trials up to and including the first success is given by $\frac{1}{p}$. This can be bounded simply by
	\[
	\frac{1}{p} = \frac{1}{1 -\frac{1}{|T_Q^n|}} \leq \frac{1}{1- \frac{1}{n}} = \frac{n}{n-1} \leq 2
	\]
	for any $n\geq 2$, using $|T_Q^n| \geq n$.

\subsection{Proof of Theorem~\ref{theo:entropic-chan-constant-type} (lower bound)}\label{sec:chan-type-achiev}

To the best of our knowledge, the achievability of such classical-quantum channel coding with fixed composition is unknown.

\begin{prop}[Achievability of Classical-Quantum Channel Coding with Fixed Composition] \label{prop:cc_ac_ch}
	For any $n\in\mathbb{N}$, $\mathscr{W}: \mathcal{X} \to \mathcal{S}(B)$, $P \in \mathscr{P}_n(\mathcal{X})$,
	there exist an $n$-blocklength channel code $\mathcal{C}$ with fixed composition $P$ and rate $R$ such that the average error probability $\Pe(\mathcal{C})$ can be bounded by
	\begin{align} \label{eq:cc_ac_ch0}
	\log \Pe(\mathcal{C}) \leq - n E_\textnormal{r,c}^\downarrow (R, P) + K \log n,
	\end{align}
	where $K$ is a constant depending on $P$ and $R$, and the entropic exponent function is defined by
	\begin{align}
	E_\textnormal{r,c}^\downarrow (R, P) :=
	\sup_{\frac12 \leq \alpha \leq 1} \frac{1-\alpha}{\alpha} \left( \sum_{x\in\mathcal{X}} P(x) D_{2-\frac{1}{\alpha}}(W_x\|P\mathscr{W}) - R \right). \label{eq:r_channel}
	\end{align}
	In particular,
	\begin{equation}
	e\chan(n,R,P) \geq E_{r,c}^{\downarrow}(R,P) - \frac{K\log n}{n}.
	\end{equation}
	
\end{prop}

To prove it, we will first prove a one-shot version given by the following proposition.

\begin{prop}[One-shot Achievability of Classical-Quantum Channel Coding] \label{prop:one-shot_ac_ch}
	For any $\mathscr{W}: \mathcal{X} \to \mathcal{S}(B)$, $P \in \mathscr{P}(\mathcal{X})$, $\mathcal{B}\subset \mathcal{X}$, and $\alpha \leq [1/2, 1]$,
	there exists a channel code $\mathcal{C}$ with codewords in $\mathcal{B}$ and $|\mathcal{C}| = M $ such that the average error probability $\Pe(\mathcal{C})$ can be bounded by
	\begin{align}
	\log \Pe(\mathcal{C}) \leq \frac{\alpha-1}{\alpha} \left[ 
	\Gamma - \log (M-1)
	+  \log \frac{P(\mathcal{B})}{6} \right]\label{eq:one-shot_ac_ch0}
	\end{align}
	where $\Gamma := \inf_{x\in\mathcal{B}} D_{2 - \frac{1}{\alpha} }\left( W_x \| P\mathscr{W} \right)$, and $P\mathscr{W} = \sum_x P(x) W_x$.
\end{prop}
\begin{proof}[Proof of Proposition~\ref{prop:one-shot_ac_ch}]
	We prove the existence of the channel codes satisfying Eq.~\eqref{eq:one-shot_ac_ch0} by using a random coding argument.
	For any $P\in\mathscr{P}(\mathcal{X})$ and $\mathcal{B} \subset \mathcal{X}$ with $P(\mathcal{B})>0$, let $P_\mathcal{B} \in \mathscr{P}(\mathcal{X})$ be 
	\begin{align}
	P_\mathcal{B} (x) := \frac{\mathds{1}_{x\in\mathcal{B}} P(x)}{ P(\mathcal{B})}.
	\end{align}
	We consider the ensemble of codes satisfying the following: the assignments of the messages $m$  to the code $\mathcal{E}(m) = x$ are jointly independent with probability $P_\mathcal{B}(x)$ for all $m$ in the message set $\mathcal{M}$.
	The decoder is characterized by the POVM $\mathcal{F} = (\Pi_m)_{m\in\mathcal{M}}$:
	\begin{align}
	\Pi_{m} &:= 	\left( \sum_{i\in\mathcal{M}} \Lambda_i  \right)^{-1/2}
	\Lambda_{ m }
	\left( \sum_{i\in\mathcal{M}} \Lambda_i  \right)^{-1/2} , \\
	\Lambda_m &:= \left\{ W_{\mathcal{E}(m)} - \gamma {P}\mathscr{W}  > 0   \right\},
	\end{align}	
	where $\gamma>0$ will be chosen later.
	Then, the average error probability of the code $\mathcal{C} = (\mathcal{E},\mathcal{F})$ is
	\begin{align}
	\Pe(\mathcal{C}) = \frac{1}{M} \sum_{m\in\mathcal{M}} \Tr\left[ W_{\mathcal{E}(m)} \left( \mathds{1} - \Pi_m \right) \right].
	\end{align}
	Invoking the Hayashi-Nagaoka inequality \cite[Lemma 2]{HN03}:
	\begin{equation}
	\mathds{1} - \Pi_{ {m} } \leq 2 \left( \mathds{1} - \Lambda_m \right) + 4 \sum_{i\neq m } \Lambda_i, \label{eq:large_ach2}
	\end{equation}
	we obtain
	\begin{align}
	\Pe(\mathcal{C}) \leq \frac{2}{M} \sum_{m\in\mathcal{M}} \Tr\left[ W_{\mathcal{E}(m)}\left\{ W_{\mathcal{E}(m)} - \gamma {P}\mathscr{W} \leq 0   \right\} \right] + \frac{4}{M} \sum_{m\in\mathcal{M}} \sum_{i\neq m} \Tr\left[ W_{\mathcal{E}(m)}\left\{ W_{\mathcal{E}(i)} - \gamma {P}\mathscr{W} > 0   \right\}
	\right].
	\end{align}
	The expected value of $\Pe(\mathcal{C})$ over the ensemble is then
	\begin{align}
	\mathbb{E}\left[ \Pe(\mathcal{C}) \right] \leq
	2 \sum_x P_\mathcal{B}(x) \Tr\left[ W_{x}\left\{ W_{x} -\gamma {P}\mathscr{W} \leq 0   \right\} \right] +
	4 (M-1) \sum_x P_\mathcal{B}(x) \Tr\left[ {P}_\mathcal{B}\mathscr{W} \left\{ W_{x} - \gamma {P}\mathscr{W} > 0   \right\} \right]. \label{eq:one-shot_ac_ch1}
	\end{align}
	
	Next, we apply Audenaert \textit{et al.}'s inequality \cite{ACM+07, ANS+08}: for every $A,B\geq 0$ and $t\in[0,1]$,
	\begin{align}
	\Tr\left[ \left\{ A - B \geq 0  \right\} B + \left\{ B - A \leq 0\right\} A \right] \leq \Tr\left[ A^t B^{1-t} \right].
	\end{align}
	Letting $A = W_x$ and $B = \gamma {P}\mathscr{W}$, the first term on the right-hand side of Eq.~\eqref{eq:one-shot_ac_ch1} can be upper bounded by
	\begin{align}
	2 \sum_x P_\mathcal{B}(x) \Tr\left[ W_{x}\left\{ W_{x} -\gamma {P}\mathscr{W} \leq 0   \right\} \right]
	&\leq 2 \sum_x P_\mathcal{B}(x) \gamma^{1-t} \Tr\left[W_x^t (P\mathscr{W})^{1-t} \right] \\
	&\leq 2  \gamma^{1-t} \exp\left\{ (t-1) \inf_{x\in\mathcal{B}} D_t(W_x\|P\mathscr{W})   \right\} \label{eq:one-shot_ac_ch2}
	\end{align}
	for all $t\in[0,1]$.
	Similarly, the second term on the right-hand side of Eq.~\eqref{eq:one-shot_ac_ch1} can be upper bounded by
	\begin{align}
	&\quad4 (M-1) \sum_x P_\mathcal{B}(x) \Tr\left[ {P}_\mathcal{B}\mathscr{W} \left\{ W_{x} - \gamma {P}\mathscr{W} > 0   \right\} \right] \notag \\
	&\leq 4 (M-1) \sum_x P_\mathcal{B}(x) \Tr\left[ {P} \mathscr{W} \left\{ W_{x} - \gamma {P}\mathscr{W} > 0   \right\} \right] \label{eq:one-shot_ac_ch3} \\
	&\leq 4 (M-1) \sum_x P_\mathcal{B}(x) \frac{1}{\gamma P(\mathcal{B})} \Tr\left[W_x^t (P\mathscr{W})^{1-t} \right] \\
	&\leq 4(M-1) \frac{1}{\gamma P(\mathcal{B})} \exp\left\{ (t-1) \inf_{x\in\mathcal{B}} D_t(W_x\|P\mathscr{W})   \right\},
	\label{eq:one-shot_ac_ch4}
	\end{align}
	where Eq.~\eqref{eq:one-shot_ac_ch3} follows from below
	\begin{align}
	P_\mathcal{B}\mathscr{W} = \sum_x \frac{\mathds{1}_{x\in\mathcal{B}} P(x) W_x }{ P(\mathcal{B}) } \leq \sum_x \frac{ P(x) W_x }{ P(\mathcal{B}) } = P\mathscr{W}.
	\end{align}
	By setting $\gamma = \frac{M-1}{P(\mathcal{B})}$, Eqs.~\eqref{eq:one-shot_ac_ch1}, \eqref{eq:one-shot_ac_ch2}, and \eqref{eq:one-shot_ac_ch4} together yield
	\begin{align}
	\mathbb{E}\left[ \Pe(\mathcal{C}) \right] &\leq \frac{6}{P(\mathcal{B})}   \exp\left\{ (t-1) \left[ \inf_{x\in\mathcal{B}} D_t(W_x\|P\mathscr{W}) - \log(M-1)  \right] \right\} \\
	&\leq \frac{\alpha-1}{\alpha} \left[ 
	\Gamma - \log (M-1)
	+  \log \frac{P(\mathcal{B})}{6} \right]
	\end{align}
	for all $\alpha\in[1/2,1]$.
	
	Sine there exists a channel coding with the average error probability less than or equal to $\mathbb{E}\left[ \Pe(\mathcal{C}) \right]$, our claim is thus proven.
	
\end{proof}

By applying Proposition~\ref{prop:one-shot_ac_ch} with the type class $T_{P}^n$ as the codeword space, we immediately arrive at the following achievability result for constant composition coding.

\begin{proof}[Proof of Proposition~\ref{prop:cc_ac_ch}]
	First note that
	\begin{align}
	P^{\otimes n} \mathscr{W}^{\otimes n} = (P\mathscr{W})^{\otimes n}.
	\end{align}
	The additivity of R\'enyi relative entropy implies that for all $\bx \in T_{P}^n$ and $\alpha\in[1/2,1]$,
	\begin{align}
	D_{2-\frac{1}{\alpha}}\left(W_{\bx}\|P^{\otimes n} \mathscr{W}^{\otimes n} \right)
	&= 	D_{2-\frac{1}{\alpha}}\left(W_{\bx}\| (P\mathscr{W})^{\otimes n} \right) \\
	&= n \sum_{x\in\mathcal{X}} P(x) D_{2-\frac{1}{\alpha}}(W_x\|P\mathscr{W}).
	\end{align}
	
	Let $\mathcal{B} = T_{P}^n$ in Proposition~\ref{prop:one-shot_ac_ch}.
	By \cite[p.~26]{CK11}, the probability of the set of all sequences with composition $P$ under the i.i.d.~distribution $P$ is 
	\begin{align}
	P^{\otimes n}\left( T_{P}^n \right) = \mathrm{e}^{- \xi \frac{|\texttt{supp}(P)|}{12 \log 2}} ( 2\pi n)^{- \frac{|\texttt{supp}(P)|-1}{2} } \sqrt{ \prod_{x:P(x)>0} \frac{1}{P(x)} }
	\end{align}
	for some $\xi\in[0,1]$.
	Hence, Proposition~\ref{prop:one-shot_ac_ch} ensures that there exists an $n$-blocklength channel code $\mathcal{C}$ with fixed composition $P$ and rate $R = \frac{\log |\mathcal{C}|}{n}$ such that
	\begin{align}
	\log \Pe(\mathcal{C}) \leq - n E_\textnormal{r,c}^\downarrow (R, P) +
	\frac{1-\alpha^\star}{\alpha^\star} \left[
	\frac{|\texttt{supp}(P)|}{12 \log 2} \frac{|\texttt{supp}(P)|-1}{2}\log(2\pi n) + \frac12 \prod_{x:P(x)>0} \frac{1}{P(x)}
	\right] \label{eq:cc_ac_ch1}
	\end{align}
	for some $\alpha^\star\in[1/2,1]$ satisfying
	\begin{align}
	R = \sum_{x\in\mathcal{X}} P(x) D_{2-\frac{1}{\alpha^\star}}(W_x\|P\mathscr{W}).
	\end{align}
	Define
	\[
	K := 2\left[\frac{|\texttt{supp}(P)|}{12 \log 2} \frac{|\texttt{supp}(P)|-1}{2}(1 +\log(2\pi)) +  \frac12 \prod_{x:P(x)>0} \frac{1}{P(x)}\right].
	\]
	Then for $n\geq 2$, $\log n \geq 1$, so
	\begin{align}
	K\log n &\geq  2\left[\frac{|\texttt{supp}(P)|}{12 \log 2} \frac{|\texttt{supp}(P)|-1}{2}( \log n +\log(2\pi)) +  \frac12 \prod_{x:P(x)>0} \frac{1}{P(x)} \right]\\
	&\geq 
	\frac{1-\alpha^\star}{\alpha^\star} \left[
	\frac{|\texttt{supp}(P)|}{12 \log 2} \frac{|\texttt{supp}(P)|-1}{2}\log(2\pi n) + \frac12 \prod_{x:P(x)>0} \frac{1}{P(x)}
	\right]
	\end{align}
	showing that the second term in the right-hand side of Eq.~\eqref{eq:cc_ac_ch1} can be upper bounded $K\log n$ for all $n>2$, which completes the proof. 
\end{proof}

\section{Strong Converse proofs with fixed type} \label{sec:sc_proofs}
\subsection{Proof of \eqref{eq:sc_LB}}
To show \eqref{eq:sc_LB}, it suffices to restrict to deterministic encoders. Let $T^n:=T^n_{Q}$.
 Fix a code $\mathcal{C} = (\mathcal{E},\mathcal{D})$ with encoder $\mathcal{E}: T^n \to \mathcal{W}^n$, decoder $\mathcal{D} = (\cD_w)_{w\in \cs}$ where $\cD_w = \{\Pi_{\bx}^{(w)}\}_{\bx\in T^n}$ is a POVM, and where $|\cs| =2^{nR}$.
	 An $n$-shot code at rate $R$ in this context is a $1$-shot code of the state $\rho^{(n)}_{T^nB^n}$ at a rate $nR$. We can thus apply the one-shot strong converse result given in the proof of Theorem 3 of [CHDH18] with a source alphabet of $T^n$: for any $\alpha > 1$ and any state $\sigma_{B^n}\in S(B^n)$,
\[
\left( 1 - \Pe(\sC)\right)^{\alpha} \left(\frac{|\cs|}{|T^n|}\right)^{1-\alpha} \leq  Q_\alpha^*\left(\rho_{T^nB^n} \left\|  \tau_{T^n} \otimes  \sigma_{B^n}\right. \right),
\]
where $\tau_{T^n} = \frac{\one_{T^n}}{|T^n|}$.
Since this holds for any state $\sigma_{B^n}\in S(B^n)$, it holds in particular for any product state $\sigma_{B^n} = \sigma_B^{\otimes n}$. Then by the joint convexity $Q_\alpha^*$ for $\alpha>1$ we have
\begin{align}	
 Q_\alpha^*\left(\rho_{T^nB^n} \left\|  \tau_{T^n} \otimes  \sigma_{B}^{\otimes n}\right. \right) &\leq \frac{1}{|T^n|} \sum_{\bx\in T^n}  Q_\alpha^*\left(\ket{\bx}\bra{\bx} \otimes \rho_{B^n}^{\bx} \left\|  \ket{\bx}\bra{\bx} \otimes \sigma_{B}^{\otimes n}\right. \right)\\
 &= \frac{1}{|T^n|} \sum_{\bx\in T^n}  Q_\alpha^*\left(\rho_{B^n}^{\bx} \left\|  \sigma_{B}^{\otimes n}\right. \right).
\end{align}
Thus,
\begin{align}	
\left( 1 - \Pe(\sC)\right)^{\alpha} \left(\frac{|\cs|}{|T^n|}\right)^{1-\alpha} 
	&\leq \frac{1}{|T^n|} \sum_{\bx\in T^n}  Q_\alpha^*\left(\rho_{B^n}^{\bx} \left\|  \sigma_{B}^{\otimes n}\right. \right).
\end{align}
By multiplicativity of $Q_\alpha^*$ under tensor products,
\begin{equation}
\frac{1}{|T^n|} \sum_{\bx\in T^n}  Q_\alpha^*\left(\rho_{B^n}^{\bx} \left\|  \sigma_{B}^{\otimes n}\right. \right) =  \frac{1}{|T^n|} \sum_{\bx\in T^n}  \prod_{x_i : \bx = (x_1,\dotsc, x_n)} Q_\alpha^*(\rho_{B}^{x_i}\|\sigma_B) = \frac{1}{|T^n|} \sum_{\bx\in T^n}  \prod_{x\in \cX} Q_\alpha^*(\rho_{B}^{x}\|\sigma_B)^{n_x}
\end{equation}
where on the far-right side we have regrouped the factors by counting the occurrences of each symbol $x\in \cX$, and
 $n_{x}$ is the number of occurrences of $x$ in the sequence $\bx$, i.e. $n_x = n Q(x)$. Since each $\bx\in T^n$ has the same type $Q$, we have
\begin{align}	
\left( 1 - \Pe(\sC)\right)^{\alpha} \left(\frac{|\cs|}{|T^n|}\right)^{1-\alpha} 
	&\leq  \prod_{x\in \cX} Q_\alpha^*(\rho_{B}^{x}\|\sigma_B)^{n Q(x)}.
\end{align}
Taking the logarithm and dividing by $n \alpha$,
\begin{gather}
	 \frac{1}{n} \log ( 1 - \Pe(\sC)) - \frac{1}{n} \frac{1-\alpha}{\alpha} \log\frac{|\cs|}{|T^n|} \leq \sum_{x\in \cX}  Q(x) \frac{1}{\alpha} \log Q_\alpha^*(\rho_{B}^{x}\|\sigma_B)\\
	 \frac{1}{n} \log ( 1 - \Pe(\sC)) + \frac{1}{n} \frac{1-\alpha}{\alpha} \log |T^n| -\frac{1-\alpha}{\alpha} R \leq \sum_{x\in \cX}  Q(x) \frac{\alpha-1}{\alpha} D_\alpha^*(\rho_{B}^{x}\|\sigma_B)\\
	 \frac{1}{n} \log ( 1 - \Pe(\sC))  \leq \frac{\alpha-1}{\alpha}  \left[\sum_{x\in \cX}  Q(x) D_\alpha^*(\rho_{B}^{x}\|\sigma_B) + R - \frac{1}{n}\log |T^n|\right]\\
	 -\frac{1}{n} \log ( 1 - \Pe(\sC))  \geq \frac{1-\alpha}{\alpha}  \left[\sum_{x\in \cX}  Q(x) D_\alpha^*(\rho_{B}^{x}\|\sigma_B) + R - \frac{1}{n}\log |T^n|\right].
\end{gather}
Since we have
\[
(n+1)^{-|\mathcal{X}|} \exp\left\{ nH(Q)  \right\} \leq |T_{Q}^n| \leq \exp\left\{ nH(Q)  \right\},
\]
\begin{align}	
 -\frac{1}{n} \log ( 1 - \Pe(\sC))  &\geq \frac{1-\alpha}{\alpha}  \left[\sum_{x\in \cX}  Q(x) D_\alpha^*(\rho_{B}^{x}\|\sigma_B) + R - H(Q) +\frac{|\cX|}{n} \log(n+1)\right] \\
 &> \frac{1-\alpha}{\alpha}  \left[\sum_{x\in \cX}  Q(x) D_\alpha^*(\rho_{B}^{x}\|\sigma_B) + R - H(Q) \right] -\frac{|\cX|}{n} \log(n+1)
\end{align}
since $1-\alpha > - \alpha \implies \frac{1-\alpha}{\alpha} > -1$. Since the left-hand side does not depend on $\sigma_B$ or $\alpha>1$ we can maximize both, and minimize over codes $\sC$, yielding
\begin{equation}
sc(n,R, Q) \geq \sup_{\alpha >1}\frac{1-\alpha}{\alpha}  \left[\inf_{\sigma_B\in S(B)}\sum_{x\in \cX}  Q(x) D_\alpha^*(\rho_{B}^{x}\|\sigma_B) + R - H(Q) \right] -\frac{|\cX|}{n} \log(n+1).
\end{equation}
Equivalently, setting $s = \frac{\alpha-1}{\alpha}$,
\begin{align}
sc(n,R, Q) &\geq \sup_{-1 < s < 0}s\left[\inf_{\sigma_B\in S(B)}\sum_{x\in \cX}  Q(x) D_{\frac{1}{1+s}}^*(\rho_{B}^{x}\|\sigma_B) + R - H(Q) \right] -\frac{|\cX|}{n} \log(n+1)\\
&= E^*_\text{sc,s}(R, Q) -\frac{|\cX|}{n} \log(n+1) \label{eq:sc_1}
\end{align}
as desired.
\subsection{Proof of \eqref{eq:sc_LB1}}
	To show \eqref{eq:sc_LB1}, it suffices to restrict to deterministic encoders. 
	Fix a code $\mathcal{C} = (\mathcal{E},\mathcal{D})$ with encoder $\mathcal{E}: \mathcal{M} \to T_{P}^n$, decoder $\mathcal{D} = \{\Pi_m\}_{m\in \mathcal{M}}$ is a POVM, and where $|\mathcal{M}| =2^{nR}$.
	Let $\sigma \in \mathcal{S}(B)$ be arbitrary and let
	\begin{align}
	X &:= \frac{1}{|\mathcal{M}|} \bigoplus_{m=1}^{|\mathcal{M}|} W_{\mathcal{E}(m)}^{\otimes n}, &
	Y &:= \frac{1}{|\mathcal{M}|} \bigoplus_{m=1}^{|\mathcal{M}|} \sigma^{\otimes n}, & 
	\Lambda &:= \bigoplus_{m=1}^{|\mathcal{M}|} \Pi_m.
	\end{align}
	Then, by the data processing inequality of $Q_\alpha^*(\cdot\|\cdot)$ for all $\alpha>1$,
	\begin{align}
	\left( 1 - \Pe(\sC)\right)^{\alpha} \frac{1}{|\mathcal{M}|^{1-\alpha}} &= 
	\Tr\left[ X\Lambda \right]^\alpha \Tr\left[ Y \Lambda \right]^{1-\alpha} \\
	&\leq \Tr\left[ X\Lambda \right]^\alpha \Tr\left[ Y \Lambda \right]^{1-\alpha} + \Tr\left[ X(1-\Lambda) \right]^\alpha \Tr\left[ Y (1-\Lambda) \right]^{1-\alpha} \\
	&\leq Q_\alpha^*(X\|Y) \\
	&= \frac{1}{|\mathcal{M}|} \sum_{m=1}^{|\mathcal{M}|} Q_\alpha^*(W_{\mathcal{E}(m)}^{\otimes n} \| \sigma^{\otimes n}  ).
	\end{align}
	
	By multiplicitivity of $Q_\alpha^*$ under tensor products,
	\begin{equation}
	\frac{1}{|\mathcal{M}|} \sum_{m=1}^{|\mathcal{M}|} Q_\alpha^*(W_{\mathcal{E}(m)}^{\otimes n} \| \sigma^{\otimes n}  ) =  \frac{1}{|\mathcal{M}|} \sum_{ m \in \mathcal{M} }  \prod_{x_i : \mathcal{E}(m) = (x_1,\dotsc, x_n)} Q_\alpha^*( W_{x_i}\|\sigma) = \frac{1}{|\mathcal{M}|} \sum_{ m \in \mathcal{M} }  \prod_{x\in \cX} Q_\alpha^*(W_{x}\|\sigma)^{n_x}
	\end{equation}
	where on the far-right side we have regrouped the factors by counting the occurences of each symbol $x\in \cX$, and
	$n_{x}$ is the number of occurences of $x$ in the sequence $\bx$, i.e. $n_x = n P(x)$. Since each $\bx\in T_P^n$ has the same type $P$, we have
	\begin{align}	
	\left( 1 - \Pe(\sC)\right)^{\alpha}  \frac{1}{|\mathcal{M}|^{1-\alpha}}
	&\leq  \prod_{x\in \cX} Q_\alpha^*( W_{x}\|\sigma)^{n P(x)}.
	\end{align}
	Taking the logarithm and dividing by $- n \alpha$,
	\begin{align}
	-\frac{1}{n} \log ( 1 - \Pe(\sC))  &\geq \frac{1-\alpha}{\alpha}  \left[\sum_{x\in \cX}  P(x) D_\alpha^*(W_{x}\|\sigma) - R \right].
	\end{align}
	Since this holds for every $\sigma\in\mathcal{S}(B)$ and $\alpha>1$, we can maximize both,
	\begin{align}
	-\frac{1}{n} \log ( 1 - \Pe(\sC))  &\geq  \sum_{\alpha>1}\frac{1-\alpha}{\alpha}  \left[ \inf_{\sigma \in \mathcal{S}(B)} \sum_{x\in \cX}  P(x) D_\alpha^*(W_{x}\|\sigma) - R \right] \\
	&= E_\text{sc,c}^*(R,P).
	\end{align}
	as desired.

\subsection{Proof of \eqref{eq:E_0_Q3_sc}}
	Recalling Eq.~\eqref{eq:prob_T}, we have
	\begin{align}
	1-\Pe(\mathcal{C}) &= \sum_{Q \in \mathscr{P}_n(\mathcal{X}) } P_{X^n} \left[x^n\in T_{Q}^n\right] (1 - \Pe(\mathcal{C},Q)) \\
	&\leq \sum_{Q \in \mathscr{P}_n(\mathcal{X}) } \exp\{ -nD(Q\|P) \} (1 - \Pe(\mathcal{C},Q))  \\
	&\leq (n+1)^{|\mathcal{X}|} \max_{Q \in \mathscr{P}(\mathcal{X})} \exp\{ -nD(Q\|P) \} (1 - \Pe(\mathcal{C},Q)).
	\label{eq:sc_2}
	\end{align}
	Combining Eqs.~\eqref{eq:sc_1}, \eqref{eq:sc_2}, and Theorem~\ref{theo:Esc-minimax-duality}, we obtain the desired result:
	\begin{align}
	sc(n,R) &\geq \min_{Q \in \mathscr{P}(\mathcal{X})} \sup_{s\in(-1,0)} \left\{ E_{0}^*(s,Q) + sR  + D(Q\|P) \right\} - 2|\mathcal{X}| \frac{\log(n+1)}{n} \\
	 &= \sup_{s\in(-1,0)} \min_{Q \in \mathscr{P}(\mathcal{X})} \left\{ E_{0}^*(s,Q) + sR  + D(Q\|P) \right\} - 2|\mathcal{X}| \frac{\log(n+1)}{n} \\
	 &= \sup_{s\in(-1,0)} \left\{ E_{0}^*(s) + sR   \right\} - 2|\mathcal{X}| \frac{\log(n+1)}{n} \\
	 &= E_\text{sc,s}^*(R) - 2|\mathcal{X}| \frac{\log(n+1)}{n}. 
	\end{align}

\printbibliography

\end{document}